\documentstyle[epsfig]{mn}
\newcommand{\cthead}[1]{\multicolumn{1}{c}{#1}}
\newcommand{\ks}{km~s$^{-1}$}
\newcommand{\kss}{km~s$^{-1}$}
\newcommand{\water}{H$_2$O}
\title[The Mopra survey] {Detection of new sources of methanol
  emission at 95~GHz with the Mopra telescope}

\author[I.E. Val'tts et al.]
{I.E. Val'tts,$^1$
 S.P. Ellingsen,$^2$
 V.I. Slysh,$^1$
 S.V. Kalenskii,$^1$
\newauthor
 R. Otrupcek$^3$ and
 G.M. Larionov$^1$\\
  $^1$Astro Space Center of Lebedev Physical Institute, Profsoyuznaya
84/32, 117810 Moscow, Russia \\
  $^2$School of Mathematics and Physics, University of Tasmania, GPO Box 252-21,
Hobart 7001, TAS, Australia \\
  $^3$Australia Telescope National Facility, PO Box 76, Epping 2121, NSW,
Australia}

\date{Received date; accepted date}
\pagerange{\pageref{firstpage}--\pageref{lastpage}}
\pubyear{1999}

\def\LaTeX{L\kern-.36em\raise.3ex\hbox{a}\kern-.15em
    T\kern-.1667em\lower.7ex\hbox{E}\kern-.125emX}

\begin{document}
\setcounter{dbltopnumber}{3}

\label{firstpage}

\maketitle

\begin{abstract}

  A southern hemisphere survey of methanol emission sources has been
  carried out using the ATNF Mopra millimetre telescope.  85 sources,
  the majority of them masers, have been detected in the $8_0-7_1A^+$
  transition of methanol at 95~GHz. Together with a similar northern
  hemisphere survey this completes the search for 95-GHz methanol
  emission from the Galactic Plane.  The previously found correlation
  between intensity of methanol emission at 44 and 95~GHz is confirmed
  here with the larger sample of sources. The results of LVG
  statistical equilibrium calculations confirm the classification of
  these sources as class~I methanol masers pumped through collisional
  excitation.

\end{abstract}

\begin{keywords}
ISM: radio lines: ISM -- masers -- surveys -- ISM: molecules
\end{keywords}

\section{Introduction}

Methanol, OH and \water\/ masers are all frequently associated with
massive star formation regions, however, methanol masers offer more
possibilities for the study of star forming regions than either OH or
\water\/, because there are numerous transitions in the microwave
region of spectrum. One of the most widespread methanol masers is the
$7_0-6_1A^+$ transition at 44~GHz.  About 50 masers from this
transition have been detected in the northern hemisphere (Morimoto,
Ohishi \& Kanzawa 1985, Haschick, Menten \& Baan 1990, Bachiller et
al. 1990, Kalenskii et al.  1992, Kalenskii et al.  1994) and a
similar number have been found in the southern hemisphere (Slysh et
al.  1994). According to the empirical scheme of Menten (1991), the 44~GHz
$7_0-6_1A^+$ transition is a class~I methanol maser.  Class~I
methanol masers differ from class~II methanol masers in that they are
not directly associated with compact H{\sc ii} regions and OH masers,
and the two classes also emit in different transitions.

The $8_0-7_1A^+$ transition at 95~GHz is an analogue of the 44-GHz
masing transition (see Fig.~1 of Val'tts et al. (1999)).  The upper
($8_0$) energy level of the 95-GHz transition is 18.5K higher than the
upper ($7_0$) energy level of the 44-GHz transition and modelling
(Cragg et al. 1992) shows that strong maser emission is also expected
from this transition.  Val'tts et al. (1995) carried out a search for
95-GHz methanol masers with the Onsala radiotelescope which detected a
large number, mostly at the position of 44-GHz masers and with spectra
similar to the spectra of the 44-GHz masers. The intensity of 95-GHz
masers was found to correlate with intensity of 44-GHz masers, and was
on average about 0.5 of the intensity 44-GHz masers. In this paper we
present the results of the first search for 95-GHz methanol masers in
the southern hemisphere, carried out with the ATNF Mopra telescope in
Australia.  This search completes a whole sky survey of class~I
methanol masers source in two transitions, at 44 and 95~GHz.

\section{Observations}

The observations were carried out in the period from July~1 to~17,
1997, using the Mopra 22-m millimeter-wave telescope of the ATNF.  The
assumed rest frequency of the $8_0-7_1A^+$ transition of methanol was
95.169489~GHz (De Lucia et al. 1989).  At this frequency only the
inner 15 metres of the Mopra antenna is illuminated and the aperture
efficiency is 41\%, which implies that one Kelvin of antenna
temperature corresponds to 40~Jy.  The half-power beamwidth of the
Mopra antenna at 95~GHz is 52$''$.  The observations were performed in
a position switching mode with reference positions offset 30$'$ in
declination.  The antenna pointing was checked and corrected every 12
hours by making observations of 86-GHz SiO masers, the nominal
pointing accuracy when this procedure is followed is 10$''$~rms.  For
the majority of sources observations were only made at the nominal
position, as time limitations did not permit the observation of grids
at offset positions (to accurately determine the position) to be made.
A grid of observations were observed toward a number of interesting
sources, and for some cases where the 95-GHz emission was anomalously
low.  A number of these grids found the strongest emission to be
offset from the nominal position and in some cases this will be due to
antenna pointing errors.  However, many of the sources have only been
observed in lower frequency class~I transitions and the offset may be
due to limited accuracy to which the source position has been
previously determined.

\begin{table*}
\begin{minipage}{140mm}
\caption{Detected 95-GHz methanol sources.  Gaussian parameters were
  determined from 64 MHz Hanning smoothed spectra, except where noted.
  a=Gaussian parameters determined from uniformly weighted spectrum;
  b=Gaussian parameters determined from 32-MHz bandwidth spectrum;
  c=Detection only marginal.
  Distances were determined using the Galactic rotation curve model
  of Brand \& Blitz (1993), except where otherwise noted.
  1=Genzel \& Stutzki (1989);
  2=Sung, Bessell \& Lee (1997);
  3=Neckel (1978);
  4=Houghton \& Whiteoak (1995);
  5=Thronson, Lowenstein \& Stokes (1979)}
\label{Detection96}
\begin{tabular}{lrrrrrrr}
\hline
\cthead{Source}&\cthead{R.A.}&\cthead{Dec.}&\cthead{Peak Flux}&\cthead{LSR radial}&\cthead{Line}&\multicolumn{2}{c}{Distance}\\
               &\cthead{1950}&\cthead{1950}&\cthead{Density}&\cthead{velocity}&\cthead{FWHM}&\cthead{Near}&\cthead{Far}\\
      &\cthead{(h m s)}&\cthead{(${}^\circ \quad {}\arcmin \quad { } \arcsec$)}&\cthead{(Jy)}&\cthead{(km s$^{-1}$)}&\cthead{(km s$^{-1}$)}&\cthead{(kpc)}&\cthead{(kpc)}\\
\hline
OMC$-$2$^{\mbox{a,b}}$
  & 05:32:59.80 & $-$05:11:29.0 &       4(2) &      10.9(0.2) &   0.9(0.5) &
 0.45$^1$ & \\
  &             &               &     138(7) &     11.4(0.01) &  0.3(0.01)&&\\
  &             &               &      85(7) &     11.5(0.01) &  0.7(0.03)&&\\
NGC2264$^{\mbox{a,b}}$
  & 06:38:24.90 &    09:32:28.0 &  99.2(5.7) &      7.3(0.01) &  0.2(0.02) &
 0.76$^2$ & \\
  &             &               &  55.2(3.7) &      7.6(0.03) &  0.7(0.04)&&\\
269.20$-$1.13$^{\mbox{a}}$
  & 09:01:52.96 & $-$48:16:07.1 &  15.1(2.3) &       9.4(0.1) &   0.7(0.1) &
 2.9 & 3.2 \\
  &             &               &   6.5(1.5) &      10.7(0.2) &   1.6(0.6)&&\\
270.26+0.84$^{\mbox{a}}$
  & 09:14:56.26 & $-$47:43:34.1 &   7.0(3.1) &       9.3(0.1) &   0.6(0.4) &
 3.1 \\
  &             &               &  21.0(1.6) &       9.8(0.1) &   2.8(0.2)&&\\
294.97$-$1.73
  & 11:36:51.54 & $-$63:12:09.4 &   5.9(0.8) &    $-$8.2(0.1) &   1.3(0.2) &
 0.6 & 6.6 \\
300.97+1.15$^{\mbox{a}}$
  & 12:32:02.60 & $-$61:23:06.0 &   9.4(1.2) &  $-$42.2(0.05) &   0.7(0.1) &
 4.2 & \\
301.14$-$0.23$^{\mbox{a}}$
  & 12:32:42.60 & $-$62:45:57.0 &  27.8(2.4) &   $-$36.3(0.2) &   1.5(0.3) &
 4.4 & \\
  &             &               &  27.2(8.1) &  $-$35.4(0.05) &   0.6(0.1)&&\\
305.21+0.21$^{\mbox{a,b}}$
  & 13:07:57.54 & $-$62:18:44.2 &  62.1(2.5) &  $-$42.4(0.01) & 0.61(0.03)&
 4.9 & \\
305.25+0.25$^{\mbox{a}}$
  & 13:08:20.50 & $-$62:16:07.0 &  13.9(0.8) &   $-$36.6(0.4) &   1.4(0.1) &
 2.5 & 7.3 \\
305.36+0.20$^{\mbox{a}}$
  & 13:09:21.00 & $-$62:17:30.0 &   7.8(1.4) &   $-$34.8(0.1) &   0.9(0.2) &
 2.8 & 7.0 \\
  &             &               &  17.4(1.4) &   $-$33.4(0.4) &   0.9(0.1)&&\\
309.39$-$0.14
  & 13:43:55.90 & $-$62:03:14.0 &   5.6(0.8) &   $-$51.0(0.3) &   5.0(0.6) &
 2.2 & 8.5 \\
  &             &               &  11.0(1.7) &   $-$50.5(0.1) &   0.7(0.1)&&\\
316.76$-$0.02$^{\mbox{a}}$
  & 14:41:07.81 & $-$59:35:33.1 &  12.9(1.6) &   $-$39.4(0.1) &   0.7(0.1) &
 2.7 & 9.7 \\
320.28$-$0.31
  & 15:06:25.71 & $-$58:14:02.2 &   5.3(1.4) &   $-$67.2(0.1) &   0.5(0.2) &
 4.6 & 8.5 \\
  &             &               &  14.5(1.9) &  $-$66.3(0.02) &   0.7(0.1)&&\\
  &             &               &   9.5(0.6) &   $-$65.1(0.2) &   2.4(0.3)&&\\
323.74$-$0.27
  & 15:27:52.41 & $-$56:20:48.1 &   4.3(1.0) &   $-$52.5(0.1) &   0.5(0.1) &
 3.3 & 10.4 \\
  &             &               &   4.9(0.4) &   $-$49.8(0.1) &   2.9(0.3)&&\\
324.72+0.34
  & 15:31:06.71 & $-$55:17:28.0 &   3.1(0.5) &   $-$50.6(0.2) &   2.5(0.4) &
 3.4 & 10.5 \\
326.475+0.703
  & 15:39:27.05 & $-$53:57:42.5 &   4.4(2.2) &   $-$41.5(0.5) &   9.6(1.9) &
 2.6 & 11.6 \\
  &             &               &  16.0(2.0) &   $-$40.9(0.1) &   3.9(0.4)&&\\
  &             &               &   7.9(1.6) &   $-$40.8(0.1) &   0.7(0.2)&&\\
326.641+0.612
  & 15:40:43.44 & $-$53:56:05.9 &  15.8(0.6) &   $-$39.4(0.1) &   4.3(0.2) &
 2.9 & 11.3 \\
327.392+0.199
  & 15:46:27.84 & $-$53:48:01.7 &   5.7(0.5) &   $-$89.4(0.1) &   1.1(0.2) &
 5.3 & 9.0 \\
  &             &               &   3.7(0.6) &   $-$88.3(0.2) &   0.8(0.3)&&\\
326.859$-$0.667
  & 15:47:20.40 & $-$54:49:03.6 &   8.9(0.7) &   $-$67.1(0.1) &   1.9(0.2) &
 3.7 & 10.5 \\
327.618$-$0.111
  & 15:48:58.76 & $-$53:54:05.4 &   6.0(0.5) &  $-$88.2(0.04) &   1.3(0.1) &
 6.5 & 7.9\\
327.29$-$0.58
  & 15:49:13.01 & $-$54:28:12.1 &   9.8(2.3) &   $-$47.2(0.4) &   3.0(0.5) &
 3.0 & 11.3 \\
  &             &               &  14.3(1.1) &   $-$44.6(0.2) &   2.5(0.9)&&\\
  &             &               &   5.1(1.6) &   $-$42.0(0.8) &   3.5(1.1)&&\\
328.81+0.64
  & 15:52:00.01 & $-$52:34:03.9 &   3.6(1.6) &   $-$43.0(0.2) &   1.6(0.6) &
 2.9 & 11.6 \\
  &             &               &   1.2(1.1) &   $-$42.0(0.2) &   0.6(0.6)&&\\
  &             &               &   1.3(0.9) &   $-$41.0(0.2) &   0.9(0.7)&&\\
  &             &               &   7.2(0.9) &   $-$40.9(0.3) &   3.1(0.4)&&\\
328.24$-$0.55
  & 15:54:06.12 & $-$53:50:47.1 &   2.7(0.6) &   $-$44.0(0.2) &   1.1(0.5) &
 2.8 & 11.6 \\
  &             &               &   4.0(0.9) &   $-$42.9(0.1) &   0.8(0.3)&&\\
  &             &               &   7.3(0.4) &   $-$41.2(0.1) &   1.8(0.2)&&\\
329.469+0.502
  & 15:55:52.44 & $-$52:14:58.0 &   3.7(0.7) &   $-$69.5(0.5) &   2.4(0.8) &
 4.5 & 10.1 \\
  &             &               &   4.5(0.6) &   $-$66.7(0.4) &   3.1(0.7)&&\\
  &             &               &   2.3(0.7) &   $-$63.4(0.1) &   0.8(0.3)&&\\
329.03$-$0.21$^{\mbox{a}}$
  & 15:56:42.01 & $-$53:04:22.1 &  19.3(1.3) &   $-$43.7(0.1) &   1.3(0.1) &
 3.1 & 11.4 \\
  &             &               &  10.3(3.3) &   $-$42.5(0.1) &   0.5(0.2)&&\\
  &             &               &  12.5(1.6) &   $-$41.7(0.1) &   0.9(0.3)&&\\
  &             &               &   9.5(1.5) &   $-$40.5(0.1) &   0.9(0.3)&&\\
329.066$-$0.308
  & 15:57:18.97 & $-$53:07:38.5 &   2.2(0.4) &   $-$42.2(0.3) &   4.3(0.9) &
 3.0 & 11.6 \\
  &             &               &   2.6(0.9) &   $-$42.0(0.1) &   0.5(0.2)&&\\
329.183$-$0.314
  & 15:57:56.13 & $-$53:03:22.4 &   7.0(0.3) &   $-$49.8(0.1) &   2.2(0.2) &
 3.6 & 11.0 \\
  &             &               &   2.5(0.5) &   $-$47.5(0.2) &   1.6(0.4)&&\\
  \hline
\end{tabular}
\end{minipage}
\end{table*}

\begin{table*}
  \begin{minipage}{140mm}
    \contcaption{}
    \label{Detection95}
    \begin{tabular}{lrrrrrrr}
      \hline
      \cthead{Source}&\cthead{R.A.}&\cthead{Dec.}&\cthead{Peak Flux}&\cthead{LSR radial}&\cthead{Line}&\multicolumn{2}{c}{Distance}\\
      &\cthead{1950}&\cthead{1950}&\cthead{Density}&\cthead{velocity}&\cthead{FWHM}&\cthead{Near}&\cthead{Far}\\
      &\cthead{(h m s)}&\cthead{(${}^\circ \quad {}\arcmin \quad { } \arcsec$)}&\cthead{(Jy)}&\cthead{(km s$^{-1}$)}&\cthead{(km s$^{-1}$)}&\cthead{(kpc)}&\cthead{(kpc)}\\
      \hline
331.13$-$0.25
  & 16:07:11.01 & $-$51:42:53.1 &  20.5(0.7) &  $-$91.0(0.02) &  1.0(0.04) &
 5.5 & 9.4 \\
  &             &               &   5.7(0.4) &   $-$88.5(0.2) &   2.8(0.5)&&\\
  &             &               &   5.7(0.4) &   $-$84.8(0.2) &   3.0(0.4)&&\\
331.442$-$0.187
  & 16:08:24.07 & $-$51:27:28.6 &   4.4(0.6) &   $-$91.5(0.1) &   0.4(0.2) &
 5.4 & 9.6\\
  &             &               &   3.9(0.6) &   $-$87.9(0.1) &   1.1(0.2)&&\\
331.34$-$0.35$^{\mbox{a}}$
  & 16:08:37.41 & $-$51:38:32.1 &  15.8(1.3) &  $-$65.7(0.03) &   0.7(0.1) &
 4.2 & 10.7 \\
332.295$-$0.094
  & 16:11:58.11 & $-$50:48:26.0 &   4.9(1.9) &   $-$49.6(0.1) &   0.6(0.2) &
 3.2 & 11.8 \\
  &             &               &   3.2(0.6) &   $-$48.8(0.3) &   1.2(0.6)&&\\
332.604$-$0.167$^{\mbox{a}}$
  & 16:13:42.20 & $-$50:38:51.4 &   8.8(0.9) &  $-$45.8(0.04) &   0.9(0.1) &
 3.5 & 11.6 \\
333.23$-$0.05$^{\mbox{a}}$
  & 16:16:02.01 & $-$50:07:50.1 &  26.3(1.0) &  $-$87.2(0.03) &   1.2(0.1) &
 5.3 & 9.9 \\
  &             &               &   2.9(1.7) &   $-$86.2(0.2) &   0.7(0.4)&&\\
333.13$-$0.43
  & 16:17:13.01 & $-$50:28:18.1 &   4.7(0.3) &   $-$55.7(0.1) &   2.2(0.3) &
 3.4 & 11.7 \\
  &             &               &   7.4(0.5) &   $-$53.1(0.1) &   1.8(0.3)&&\\
  &             &               &  4.17(0.6) &   $-$51.4(0.2) &   0.9(0.6)&&\\
  &             &               &  4.30(1.1) &   $-$50.6(0.1) &   0.5(0.1)&&\\
  &             &               &  11.3(0.4) &   $-$48.5(0.1) &   3.1(0.2)&&\\
333.61$-$0.22$^{\mbox{c}}$
  & 16:18:25.01 & $-$49:59:08.1 &   3.0(0.4) &   $-$49.7(0.2) &   2.9(0.4) &
 3.4 & 11.8 \\
335.59$-$0.29$^{\mbox{a}}$
  & 16:27:15.40 & $-$48:37:20.1 &   6.8(2.3) &   $-$47.5(0.1) &   0.9(0.3) &
 3.3 & 12.2 \\
  &             &               &  10.2(2.4) &   $-$46.6(0.2) &   4.7(0.6)&&\\
  &             &               &   6.8(2.6) &   $-$46.3(0.1) &   0.5(0.2)&&\\
  &             &               &  22.3(2.2) &  $-$45.5(0.03) &   0.6(0.1)&&\\
336.41$-$0.26
  & 16:30:31.51 & $-$48:00:08.1 &   6.6(0.7) &   $-$89.7(0.1) &   1.4(0.2) &
 5.3 & 10.3 \\
  &             &               &  24.8(0.9) &  $-$87.6(0.01) &  0.7(0.03)&&\\
337.40$-$0.41$^{\mbox{a}}$
  & 16:35:08.01 & $-$47:22:23.1 &   9.4(2.1) &   $-$44.2(0.1) &   0.4(0.1) &
 3.3 & 12.4 \\
  &             &               &  16.5(2.4) &   $-$42.3(0.1) &   0.2(0.1)&&\\
  &             &               &   5.7(0.5) &   $-$40.7(0.3) &   7.4(0.8)&&\\
338.92+0.56$^{\mbox{a}}$
  & 16:36:54.01 & $-$45:36:05.0 &  18.4(1.8) &   $-$62.9(0.1) &   1.1(0.1) &
 4.4 & 11.5 \\
  &             &               &   8.3(1.4) &   $-$60.1(0.2) &   1.5(0.4)&&\\
337.91$-$0.47$^{\mbox{a}}$
  & 16:37:27.01 & $-$47:02:10.1 &  58.2(2.4) &  $-$43.6(0.01) &  0.4(0.02) &
 3.3 & 12.4 \\
341.19$-$0.22
  & 16:48:39.50 & $-$44:23:33.5 &   6.1(0.9) &   $-$42.0(0.1) &   0.7(0.1) &
 3.5 & 12.6 \\
341.22$-$0.21$^{\mbox{a}}$
  & 16:48:42.10 & $-$44:21:53.0 &  15.6(1.0) &   $-$43.8(0.1) &   1.8(0.1) &
 3.5 & 12.5 \\
  &             &               &   6.4(2.2) &   $-$43.0(0.1) &   0.4(0.2)&&\\
345.01+1.79$^{\mbox{a,b}}$
  & 16:53:21.00 & $-$40:09:40.0 &  25.6(1.1) &  $-$13.6(0.04) &   2.8(0.1) &
 1.6 & 14.8 \\
  &             &               &  42.0(1.9) &  $-$13.1(0.01) &  0.5(0.03)&&\\
343.12$-$0.06$^{\mbox{a}}$
  & 16:54:43.00 & $-$42:47:49.0 &  46.6(3.0) &  $-$32.8(0.02) &   0.8(0.1) &
 3.1 & 13.2 \\
  &             &               &  24.3(2.9) &  $-$31.5(0.02) &   0.5(0.1)&&\\
  &             &               &  22.1(1.9) &   $-$31.2(0.2) &   2.4(0.2)&&\\
  &             &               &  19.7(1.5) &  $-$27.3(0.02) &   0.7(0.1)&&\\
344.23$-$0.57$^{\mbox{a}}$
  & 17:00:35.16 & $-$42:14:29.7 &  12.0(0.8) &   $-$21.7(0.1) &   3.9(0.3) &
 2.2 & 14.1 \\
  &             &               &  33.2(1.9) &  $-$20.3(0.02) &   0.6(0.1)&&\\
345.51+0.35
  & 17:00:54.00 & $-$40:40:02.0 &   4.6(0.3) &   $-$17.1(0.1) &   4.1(0.2) &
 2.0 & 14.4 \\
  &             &               &   3.2(0.5) &   $-$16.5(0.1) &   0.9(0.2)&&\\
  &             &               &   2.3(0.6) &   $-$14.7(0.1) &   0.5(0.1)&&\\
345.00$-$0.22
  & 17:01:38.51 & $-$41:24:59.0 &  10.0(1.2) &  $-$28.9(0.03) &   0.6(0.1) &
 3.0 & 13.5 \\
  &             &               &  10.0(1.2) &   $-$28.0(0.1) &   3.2(0.4)&&\\
  &             &               &  23.5(1.4) &  $-$28.0(0.02) &   0.7(0.1)&&\\
  &             &               &   3.4(1.1) &   $-$26.0(0.6) &  10.2(1.7)&&\\
  &             &               &  10.6(0.8) &  $-$23.9(0.03) &   1.2(0.1)&&\\
349.10+0.11
  & 17:13:01.00 & $-$37:56:06.0 &   6.4(1.6) &   $-$78.0(0.1) &   0.5(0.2) &
 6.0 & 10.7 \\
  &             &               &   4.4(0.9) &   $-$77.3(0.2) &   2.3(0.4)&&\\
351.16+0.70$^{\mbox{a}}$
  & 17:16:35.51 & $-$35:54:44.0 &  22.2(5.9) &    $-$7.2(0.5) &   2.4(0.6) &
 1.4 & 15.4 \\
  &             &               &   8.0(3.7) &    $-$7.1(0.1) &   0.3(0.2)&&\\
  &             &               &  11.8(8.6) &    $-$5.8(0.3) &   1.1(0.5)&&\\
  &             &               &  26.3(3.3) &   $-$4.9(0.03) &   0.6(0.1)&&\\
351.24+0.67
  & 17:16:54.51 & $-$35:51:58.0 &   4.3(0.4) &    $-$4.5(0.1) &   3.4(0.3) &
 0.1 & 16.1 \\
351.41+0.64$^{\mbox{a}}$
  & 17:17:32.35 & $-$35:44:04.2 &  20.2(1.0) &    $-$8.2(0.2) &   4.2(0.3) &
 1.7$^3$ & \\
  &             &               &   7.0(1.3) &    $-$6.4(0.1) &   0.7(0.2)&&\\
  &             &               &   6.6(1.6) &    $-$4.6(0.5) &   3.6(0.8)&&\\
  \hline
\end{tabular}
\end{minipage}
\end{table*}

\begin{table*}
  \begin{minipage}{140mm}
    \contcaption{}
    \begin{tabular}{lrrrrrrr}
      \hline
      \cthead{Source}&\cthead{R.A.}&\cthead{Dec.}&\cthead{Peak Flux}&\cthead{LSR radial}&\cthead{Line}&\multicolumn{2}{c}{Distance}\\
      &\cthead{1950}&\cthead{1950}&\cthead{Density}&\cthead{velocity}&\cthead{FWHM}&\cthead{Near}&\cthead{Far}\\
      &\cthead{(h m s)}&\cthead{(${}^\circ \quad {}\arcmin \quad { } \arcsec$)}&\cthead{(Jy)}&\cthead{(km s$^{-1}$)}&\cthead{(km s$^{-1}$)}&\cthead{(kpc)}&\cthead{(kpc)}\\
      \hline
NGC6334I(N)$^{\mbox{a}}$
  & 17:17:33.00 & $-$35:42:04.0 &  42.6(4.0) &   $-$7.5(0.03) &   0.9(0.1) &
 1.7$^3$ & \\
  &             &               &  75.9(7.0) &   $-$4.8(0.04) &   1.7(0.2)&&\\
  &             &               &  77.9(5.7) &   $-$4.8(0.01) &  0.3(0.03)&&\\
  &             &               &  50.2(6.4) &    $-$4.2(0.2) &   5.4(0.3)&&\\
  &             &               &  52.7(4.1) &   $-$3.0(0.03) &   0.8(0.1)&&\\
351.78$-$0.54$^{\mbox{a}}$
  & 17:23:20.67 & $-$36:06:45.4 &  23.0(1.9) &   $-$8.3(0.02) &   0.5(0.1) &
 1.5 & 15.3 \\
  &             &               &  36.0(2.2) &   $-$7.5(0.03) &   0.5(0.1)&&\\
  &             &               &  73.0(2.3) &   $-$6.9(0.01) &  0.4(0.03)&&\\
  &             &               &  26.8(1.2) &    $-$3.9(0.1) &   7.1(0.2)&&\\
  &             &               &  51.6(2.3) &   $-$2.4(0.01) &  0.4(0.02)&&\\
  &             &               &  22.2(1.6) &    $-$2.0(0.1) &   2.4(0.2)&&\\
351.64$-$1.26
  & 17:25:55.00 & $-$36:37:48.0 &  10.2(1.2) &   $-$13.0(0.1) &   1.2(0.2) &
 2.5 & 14.3 \\
354.61+0.47$^{\mbox{a}}$
  & 17:27:00.00 & $-$33:11:38.0 &   9.6(2.1) &  $-$21.4(0.04) &   0.4(0.1) &
 4.0 & 12.9 \\
  &             &               &   7.8(0.8) &   $-$20.5(0.2) &   3.0(0.3)&&\\
  &             &               &  24.0(1.7) &  $-$17.9(0.02) &  0.5(0.04)&&\\
353.41$-$0.36
  & 17:27:07.00 & $-$34:39:41.0 &   6.5(0.4) &   $-$17.4(0.1) &   5.0(0.3) &
 3.4 & 13.5 \\
  &             &               &   8.1(0.8) &  $-$16.6(0.03) &   0.6(0.1)&&\\
SgrA$-$F$^{\mbox{a}}$
  & 17:42:27.40 & $-$29:02:18.0 &   8.9(0.4) &      17.8(0.7) &   2.1(1.1)&&\\
  &             &               &   4.1(1.2) &      23.9(0.3) &   4.2(1.1)&&\\
359.62$-$0.25
  & 17:42:30.00 & $-$29:22:31.0 &   5.3(0.7) &      17.7(0.1) &   1.7(0.3)&&\\
  &             &               &  15.9(1.1) &     19.3(0.03) &   0.6(0.1)&&\\
  &             &               &   9.0(0.8) &      20.6(0.1) &   1.2(0.1)&&\\
Sgr A$-$A$^{\mbox{a}}$
  & 17:42:41.30 & $-$28:58:18.0 &  12.1(0.2) &      42.1(0.2) &  18.3(0.4)&&\\
Sgr B2$^{\mbox{a}}$
  & 17:44:10.60 & $-$28:22:05.0 &  34.1(4.3) &      59.0(1.9) &  14.4(4.2) &
 8.5$^4$ &\\
  &             &               &   7.8(0.3) &      70.5(3.4) &   8.2(9.0)&&\\
0.54$-$0.85
  & 17:47:04.10 & $-$28:54:01.0 &   4.1(0.9) &      15.2(0.2) &   2.3(0.4) &
 7.5 & 9.5 \\
  &             &               &   4.0(0.9) &      16.9(0.1) &   0.6(0.2)&&\\
  &             &               &   4.7(0.3) &      18.3(0.4) &   4.1(0.7)&&\\
5.89$-$0.39$^{\mbox{a}}$
  & 17:57:26.80 & $-$24:03:54.0 &   9.5(0.8) &       8.5(0.1) &   1.2(0.1) &
 2.4 & 14.5 \\
  &             &               &   6.1(0.5) &       9.6(0.2) &   7.7(0.5)&&\\
  &             &               &   9.2(1.3) &     11.6(0.03) &   0.4(0.1)&&\\
  &             &               &   9.7(2.5) &     12.3(0.03) &   0.3(0.1)&&\\
M8E$^{\mbox{a,b}}$
  & 18:01:49.71 & $-$24:26:56.0 & 129.7(2.0) &     11.0(0.01) &  0.5(0.01) &
 1.78$^5$\\
8.67$-$0.36
  & 18:03:19.00 & $-$21:37:59.0 &   4.9(0.7) &      33.5(0.3) &   1.5(0.5) &
 4.5 & 12.3\\
  &             &               &   5.9(1.2) &      35.0(0.2) &   1.2(0.4)&&\\
  &             &               &   4.3(1.0) &      36.2(0.1) &   0.6(0.2)&&\\
IRAS18056$-$1952
  & 18:05:38.00 & $-$19:52:34.0 &   9.5(0.9) &     61.7(0.02) &   0.4(0.1) &
 5.8 & 10.9 \\
10.6$-$0.4
  & 18:07:30.50 & $-$19:56:28.0 &   8.1(0.9) &    $-$7.6(0.3) &   2.2(0.4) &
 2.0 & 18.7 \\
  &             &               &   7.3(2.2) &    $-$6.4(0.1) &   1.2(0.2)&&\\
  &             &               &   3.9(0.6) &    $-$4.7(0.1) &   1.0(0.2)&&\\
  &             &               &   4.1(0.5) &    $-$2.8(0.1) &   1.7(0.3)&&\\
  &             &               &   3.2(0.3) &    $-$1.4(0.7) &   9.6(0.9)&&\\
12.89+0.49$^{\mbox{b}}$
  & 18:08:56.40 & $-$17:32:14.0 &   6.0(1.2) &     29.6(0.04) &   0.4(0.1) &
 3.5 & 13.1 \\
  &             &               &   5.7(1.1) &      31.5(0.1) &   0.4(0.2)&&\\
  &             &               &   5.1(0.4) &      32.9(0.2) &   3.3(0.4)&&\\
11.94$-$0.62
  & 18:11:04.40 & $-$18:54:20.0 &   3.1(0.5) &      35.0(0.1) &   1.4(0.2) &
 3.9 & 12.7 \\
  &             &               &   2.7(0.6) &      36.8(0.1) &   0.6(0.2)&&\\
  &             &               &   5.5(0.4) &      38.5(0.1) &   1.3(0.1)&&\\
W33MetC$^{\mbox{a}}$
  & 18:11:15.70 & $-$17:56:53.0 &  21.2(3.1) &     32.7(0.04) &   0.5(0.1) &
 3.4 & 13.2 \\
IRAS18141$-$1615
  & 18:14:09.00 & $-$16:15:47.0 &   4.0(1.1) &      21.9(0.1) &   0.8(0.3) &
 2.1 & 14.4  \\
  &             &               &   3.7(0.7) &      23.1(0.3) &   0.8(1.0)&&\\
  &             &               &   7.2(1.0) &      25.0(0.2) &   2.1(0.4)&&\\
IRAS18151$-$1208
  & 18:15:09.00 & $-$12:08:34.0 &   3.3(0.4) &      32.2(0.1) &   1.3(0.2) &
 2.8 & 13.4 \\
  &             &               &   3.9(0.7) &      33.9(0.1) &   0.6(0.1)&&\\
14.33$-$0.64$^{\mbox{a}}$
  & 18:16:00.80 & $-$16:49:06.0 &  38.9(1.8) &     19.1(0.02) &   0.9(0.1) &
 2.6 & 13.9 \\
  &             &               &  32.2(2.0) &     20.4(0.03) &   0.7(0.1)&&\\
  &             &               &  11.8(2.3) &      21.3(0.1) &   0.7(0.2)&&\\
  &             &               &  50.1(1.8) &     22.5(0.02) &   1.0(0.1)&&\\
  &             &               &  52.8(2.9) &     23.4(0.02) &  0.5(0.04)&&\\
  \hline
\end{tabular}
\end{minipage}
\end{table*}

\begin{table*}
  \begin{minipage}{140mm}
    \contcaption{}
    \begin{tabular}{lrrrrrrr}
      \hline
      \cthead{Source}&\cthead{R.A.}&\cthead{Dec.}&\cthead{Peak Flux}&\cthead{LSR radial}&\cthead{Line}&\multicolumn{2}{c}{Distance}\\
      &\cthead{1950}&\cthead{1950}&\cthead{Density}&\cthead{velocity}&\cthead{FWHM}&\cthead{Near}&\cthead{Far}\\
      &\cthead{(h m s)}&\cthead{(${}^\circ \quad {}\arcmin \quad { } \arcsec$)}&\cthead{(Jy)}&\cthead{(km s$^{-1}$)}&\cthead{(km s$^{-1}$)}&\cthead{(kpc)}&\cthead{(kpc)}\\
      \hline
GGD 27$^{\mbox{a,b}}$
  & 18:16:13.80 & $-$20:48:31.0 &  19.3(2.7) &     12.4(0.02) &   0.3(0.1) &
 2.1 & 14.6 \\
  &             &               &  33.4(3.9) &     12.9(0.02) &  0.3(0.04)&&\\
  &             &               &  76.2(2.8) &     13.3(0.01) &  0.4(0.03)&&\\
  &             &               &  14.8(3.2) &     13.7(0.02) &   0.2(0.1)&&\\
M17(3)
  & 18:17:31.00 & $-$16:12:50.0 &   1.9(1.9) &      19.0(1.2) &   2.0(1.6) &
 2.2 & 14.2 \\
  &             &               &   2.8(3.0) &      19.1(0.1) &   0.4(0.2)&&\\
  &             &               &   2.8(0.5) &      19.4(0.3) &   5.4(0.7)&&\\
16.59$-$0.06
  & 18:18:20.30 & $-$14:33:18.0 &   5.7(1.1) &      60.4(0.2) &   1.1(0.4) &
 4.7 & 11.6 \\
  &             &               &   9.8(1.6) &      61.3(0.1) &   0.8(0.2)&&\\
19.61$-$0.23$^{\mbox{a}}$
  & 18:24:50.30 & $-$11:58:34.0 &   6.2(0.7) &      41.3(0.1) &   1.9(0.3) &
 3.4 & 12.6 \\
L379IRS3
  & 18:26:32.90 & $-$15:17:58.0 &   3.9(0.8) &      17.8(0.3) &   9.2(1.2) &
 1.7 & 14.6 \\
  &             &               &   4.6(1.3) &      17.7(0.1) &   0.7(0.2)&&\\
  &             &               &  10.1(1.2) &      18.0(0.1) &   2.7(0.3)&&\\
  &             &               &   6.0(0.8) &      20.3(0.1) &   0.9(0.2)&&\\
23.43$-$0.19$^{\mbox{a}}$
  & 18:31:55.80 & $-$08:34:17.0 &   7.0(1.4) &      96.6(0.2) &   1.7(0.4) &
 5.8 & 9.8 \\
  &             &               &  33.8(2.5) &     99.6(0.02) &  0.5(0.04)&&\\
23.01$-$0.41$^{\mbox{c}}$
  & 18:31:56.70 & $-$09:03:18.0 &   3.4(0.6) &      77.4(0.1) &   1.3(0.3) &
 4.9 & 10.8 \\
29.96$-$0.02
  & 18:43:27.10 & $-$02:42:36.0 &   3.3(0.6) &      96.2(0.1) &   0.7(0.2) &
 6.1 & 8.6 \\
  &             &               &   3.8(0.6) &      97.1(0.1) &   0.7(0.2)&&\\
  &             &               &   3.2(0.5) &      98.7(0.1) &   1.1(0.2)&&\\
  &             &               &   2.8(0.7) &     100.4(0.1) &   0.9(0.2)&&\\
30.69$-$0.06
  & 18:44:58.90 & $-$02:04:27.0 &   3.2(0.3) &      91.4(0.2) &   5.0(0.4) &
 5.3 & 9.3 \\
  &             &               &   3.8(0.7) &      93.1(0.1) &   0.6(0.2)&&\\
IRAS18537+0749$^{\mbox{a}}$
  & 18:53:46.00 & 07:49:16.0    &   5.9(1.1) &      30.0(0.3) &   1.6(0.6) &
 2.2 & 10.8 \\
  &             &               &  26.7(2.2) &      31.5(0.1) &   0.9(0.2)&&\\
  &             &               &  22.3(5.0) &      32.3(0.1) &   0.6(0.2)&&\\
  &             &               &  18.4(1.3) &      33.3(0.1) &   1.5(0.2)&&\\
35.05$-$0.52$^{\mbox{a}}$
  & 18:54:37.10 & 01:35:01.0    &  16.1(1.1) &     49.9(0.02) &   0.7(0.1) &
 3.3 & 10.6 \\
W51e1/e2
  & 19:21:26.20 & 14:54:43.0    &  17.7(0.5) &      55.9(0.1) &   5.3(0.2) &
 5.4 &\\
  &             &               &   4.8(0.7) &      61.1(0.2) &   2.6(0.5)&&\\
W51Met2$^{\mbox{a,b}}$
  & 19:21:28.90 & 14:23:48.0    &  10.5(2.0) &      54.2(0.1) &   1.0(0.2) &
 5.5 & \\
  &             &               &  19.7(2.6) &     56.7(0.04) &   0.6(0.1)&&\\
    \hline
\end{tabular}
\end{minipage}
\end{table*}

A cryogenically cooled low-noise SIS~mixer was used in the receiver.
The single side-band receiver noise temperature was 110~K and the
system temperature varied between 220~K and 320~K depending on weather
conditions and the elevation of the telescope.  An ambient temperature
load (assumed to have a temperature of 290~K) was regularly placed in
front of the receiver to enable calibration using the method of Kutner
\& Ulich (1981).  This corrects the observed flux density for the
effects of atmospheric absorption, ohmic losses and rear-ward
spillover.  Variations in the ambient temperature of a few percent
occurred during the observations, and the estimated uncertainty of the
absolute flux density scale is 10\%.

For the majority of the observations the back-end was a 64~MHz wide
1024-channel autocorrelator with a frequency resolution of 62.5~kHz.
This yields a velocity resolution at 95~GHz of 0.236~\kss with uniform
weighting and 0.394~\kss with Hanning smoothing.  Some sources were
reobserved with the correlator configured to a 32~MHz bandwidth, also
with 1024 spectral channels.  This yields a velocity resolution of
0.118~\kss with uniform weighting and 0.197~\kss with Hanning
smoothing.  For each source a uniformly weighted spectrum was produced
with a velocity width of approximately 80~\kss centred on the velocity
of the previously detected class~I or II methanol maser emission.  The
spectrum was then Hanning smoothed to improve the signal-to-noise
ratio of weak sources.  For the Mopra observations, the spectra in all
figures and Gaussian parameters (peak flux density, velocity and full
width half maximum) in all tables are Hanning smoothed data collected
with the 64-MHz correlator configuration, unless otherwise noted in
Table~\ref{Detection96}.

The source list was compiled primarilty from 44-GHz class~I methanol
masers detected in the southern hemisphere by Slysh et al. (1994) and
Haschick et al. (1990). The observing list also included 12 class~I
methanol masers detected at 36~GHz at Puschino (Kalenskii, private
communication), 55 class~II methanol sources which had not previously
been searched for class~I maser emission (Caswell et al. 1995,
Ellingsen et al. 1996, Ellingsen et al. 1999, Walsh et al. 1997) and
12 H{\sc ii} regions for which no 44-GHz emission was detected by
Slysh et al. (1994).

\section{Results}

\begin{figure*}
\resizebox{1.05\textwidth}{1.15\height}{\includegraphics{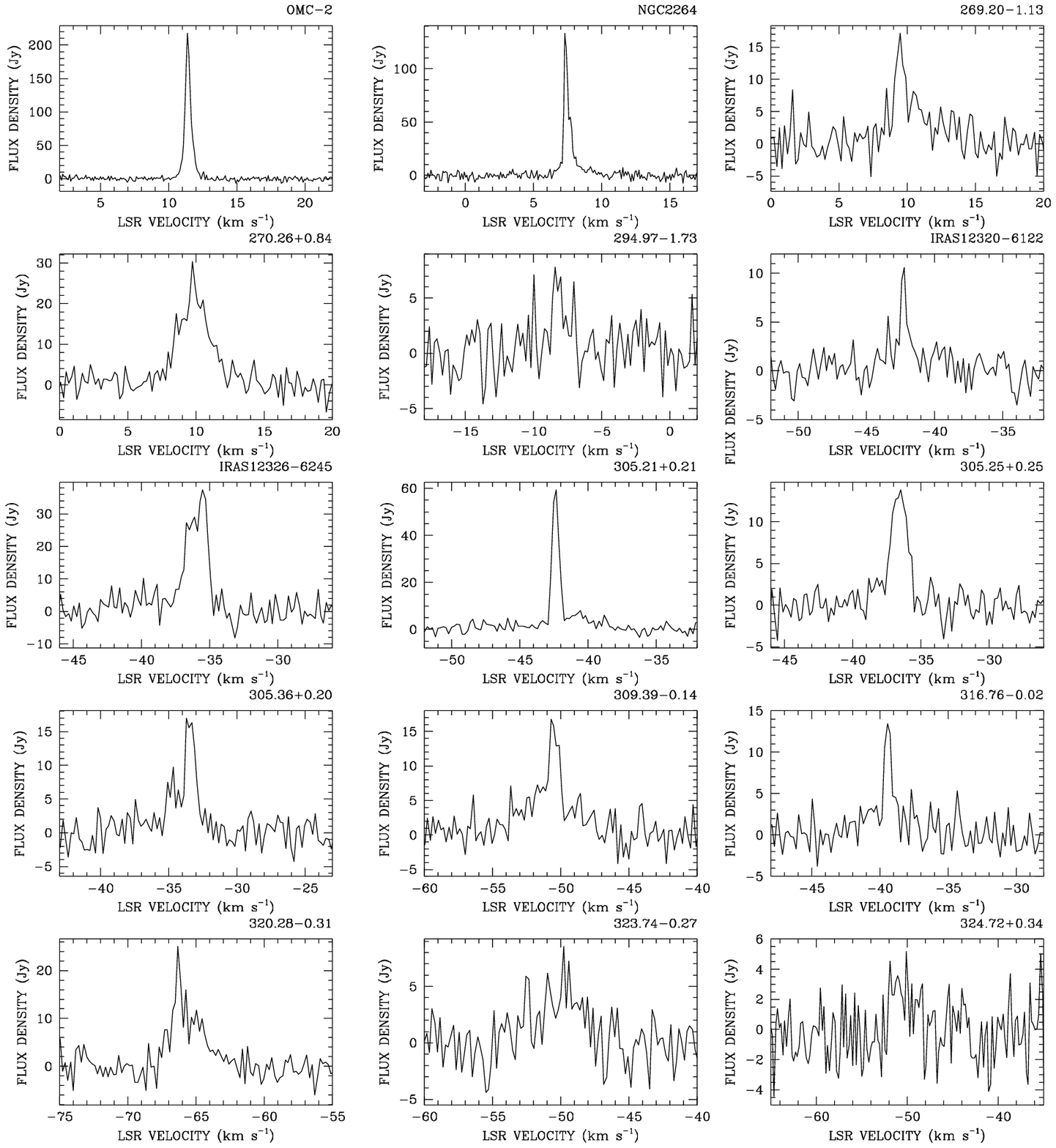}}
\vskip -5mm
\caption{95-GHz spectra.  All spectra are Hanning smoothed.}
\label{Spectra95}
\end{figure*}

\begin{figure*}
\resizebox{1.05\textwidth}{1.15\height}{\includegraphics{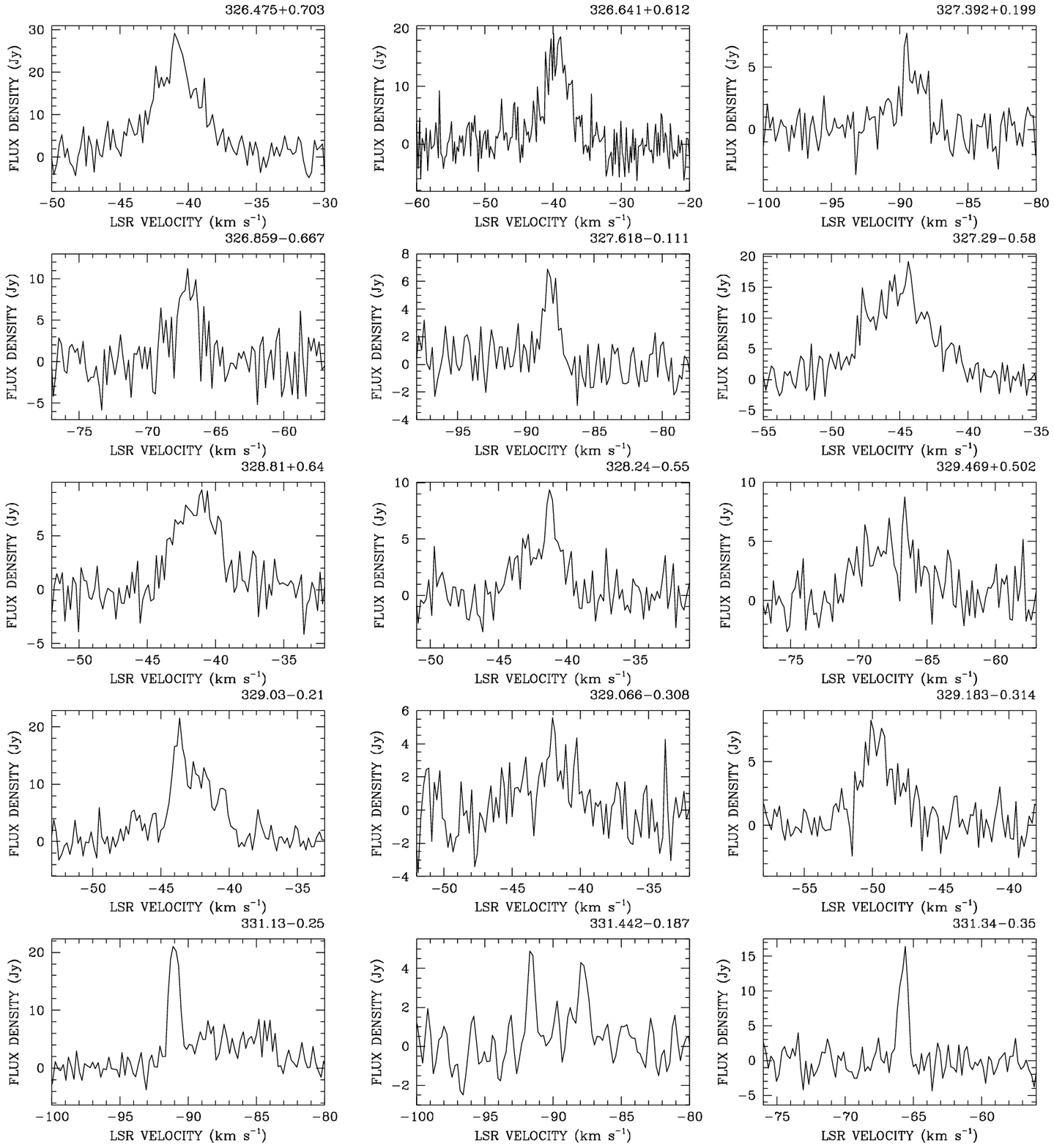}}
\vskip -5mm
\contcaption{}
\end{figure*}

\begin{figure*}
\resizebox{1.05\textwidth}{1.15\height}{\includegraphics{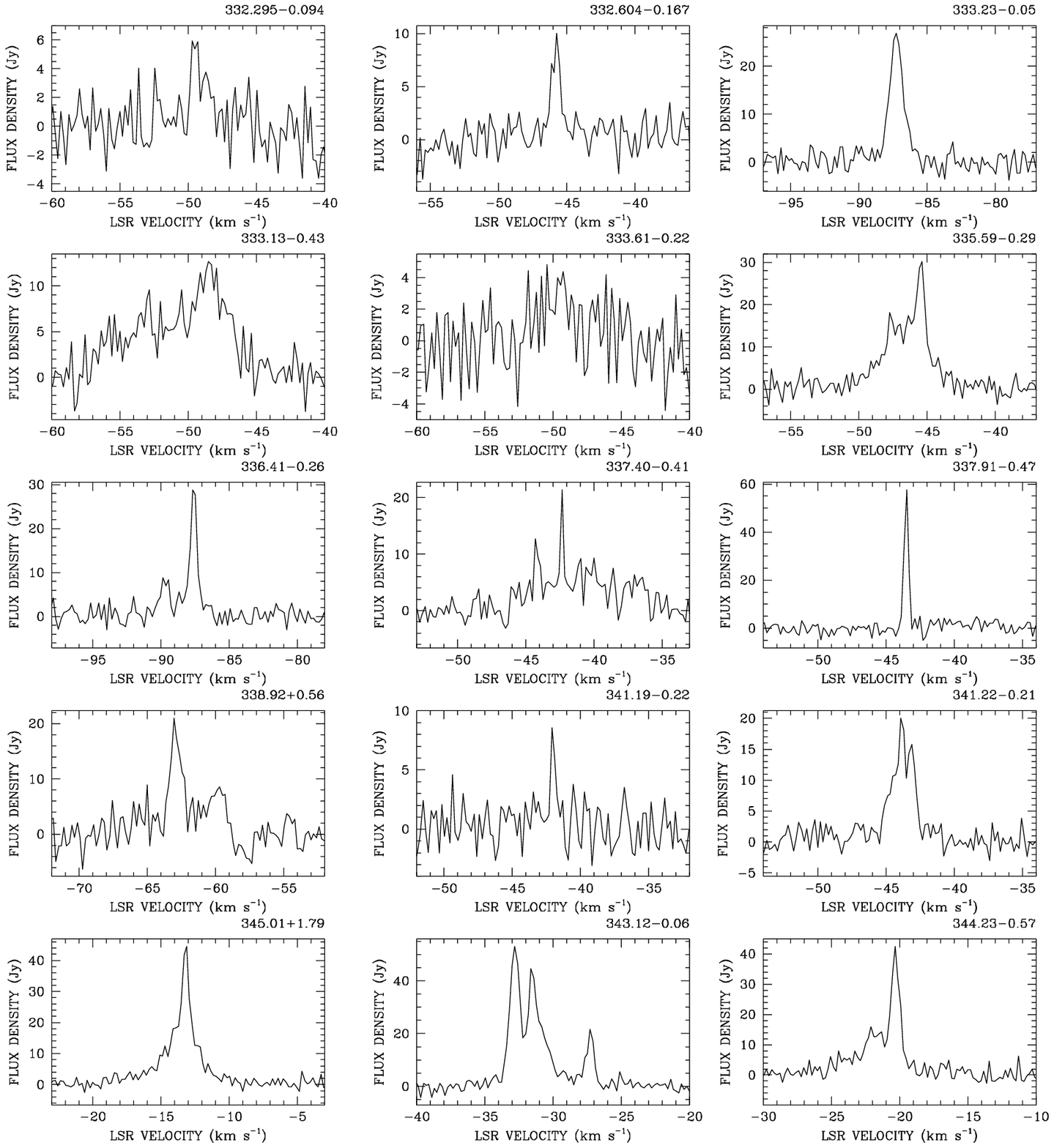}}
\vskip -5mm
\contcaption{}
\end{figure*}

\begin{figure*}
\resizebox{1.05\textwidth}{1.15\height}{\includegraphics{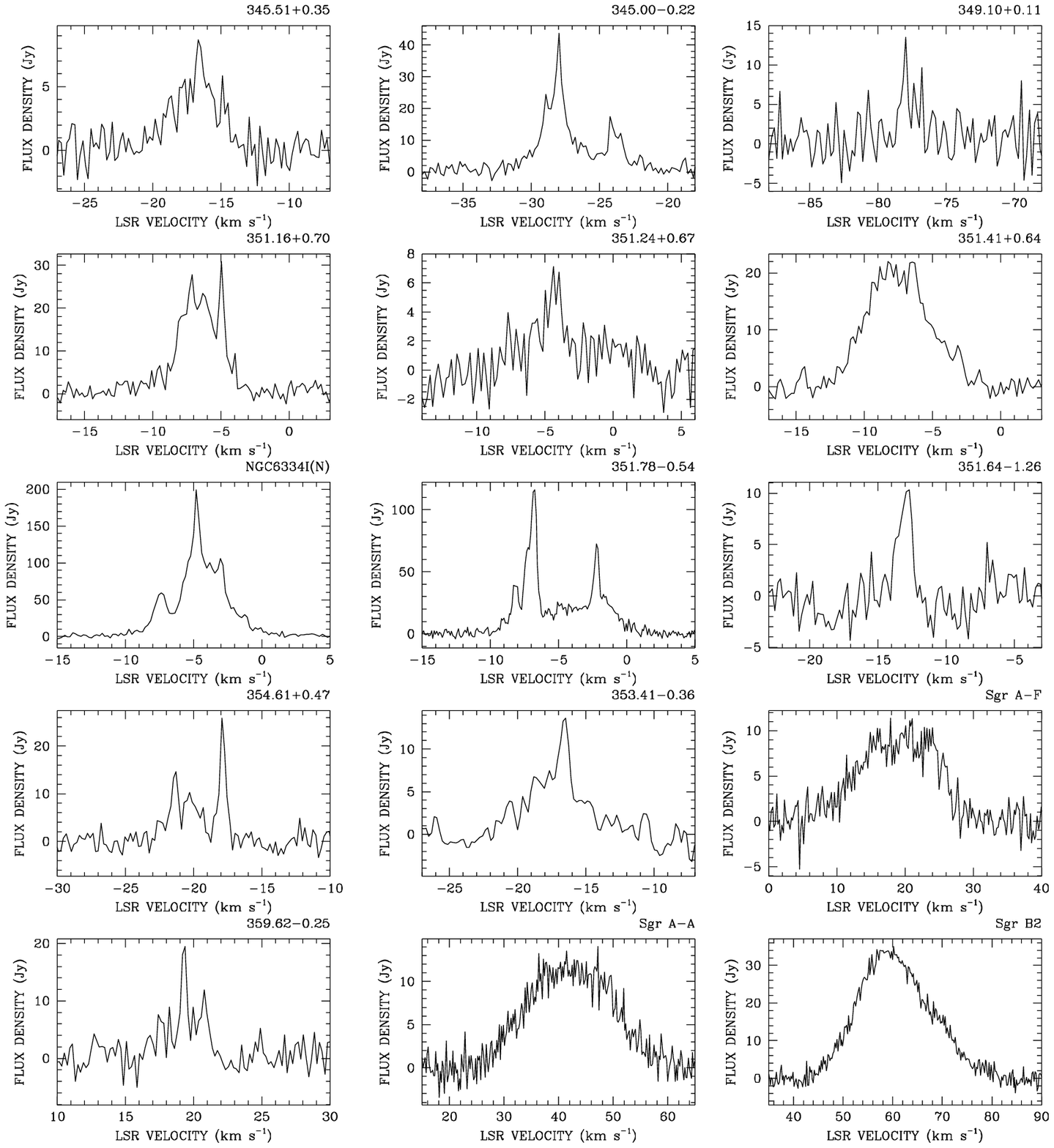}}
\vskip -5mm
\contcaption{}
\end{figure*}

\begin{figure*}
\resizebox{1.05\textwidth}{1.15\height}{\includegraphics{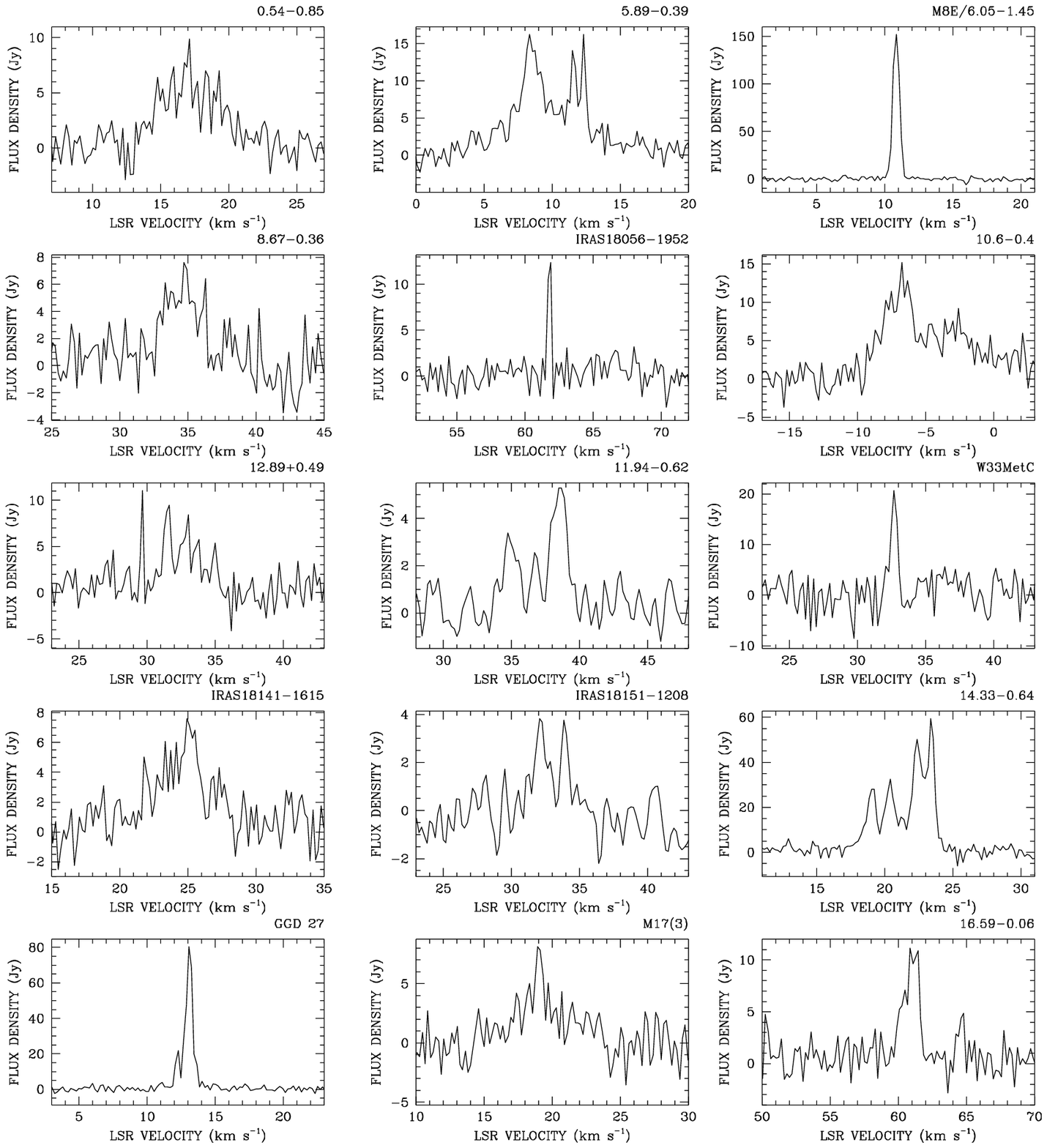}}
\vskip -5mm
\contcaption{}
\end{figure*}

\begin{figure*}
\resizebox{1.05\textwidth}{1.15\height}{\includegraphics{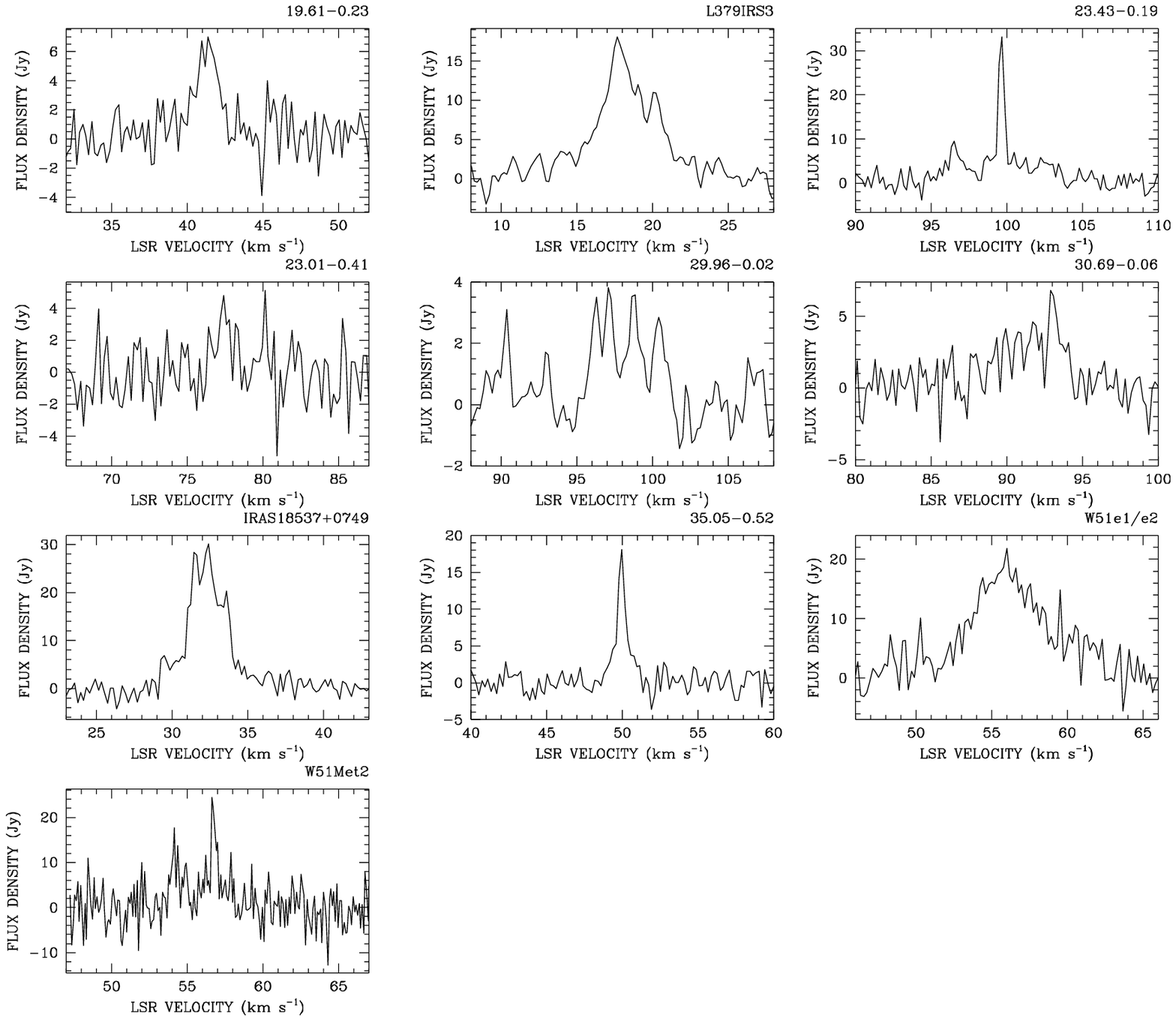}}
\vskip -5mm
\contcaption{}
\end{figure*}

A total 153 sites were searched for 95-GHz methanol emission, with
detections observed toward 85.  Table~\ref{Detection96} contains a
list of all detected sources along with the Gaussian parameters of
their spectral features.  Their spectra are shown in Fig.~1.  The
sources toward which no 95-GHz emission was detected are listed in
Table~\ref{Negative95}, along with the 3$-\sigma$ limit (typically
less than 5~Jy).  The kinematic distance to each source was estimated
using the rotation curve of Brand \& Blitz (1993).  For many of the
sources the model yields two distance estimates, The majority are
likely to be at the nearer distance, but we have not attempted to
resolve the ambiguity and where it occurs both distances are listed in
Table~\ref{Detection96} \& \ref{Negative95}.  86 known 44- or 36-GHz
class I methanol sources were observed, with 95-GHz emission being
detected toward 71 (83\%).  A total of 54 class~II methanol maser
sites which had not previously been searched for class I methanol
emission were also observed, yielding a total of 13 detections (24\%).

Seventy nine of the 85 sources in Table~\ref{Detection96} are new
detections, the other 6 were discovered previously by Val'tts et al.
(1995) at Onsala.  In Table~\ref{Onsala_Mopra95} we compare the
Gaussian parameters of the strongest features of 6 sources observed at
both Onsala and Mopra.  The radial velocities agree within 0.4~\ks in
all cases, which is approximately half the spectral resolution of the
Onsala observations.  The flux densities agree to within a factor of
2, which is reasonable, taking into account differences in the
spectral resolution, absolute calibration and possible pointing
errors.  Examination of Table~\ref{Onsala_Mopra95} shows that some of
the masers (e.g. OMC-2 and NGC2264) have very narrow lines of 0.2~\ks,
which were completely unresolved with the 0.7~\ks spectral resolution
of the Onsala observations. The Onsala results have not been corrected
for spectral smoothing, and for this reason, very narrow features in
OMC-2, NGC2264 and W51Met2 will have lower peak flux density than that
observed at Mopra.  Below we give comments for some of the more
interesting sources.

\begin{table*}
\begin{minipage}{140mm}
\caption{Sources undetected at 95~GHz.}
\label{Negative95}
\begin{tabular}{lrrrrrr}
\hline
\cthead{Source}  & \cthead{R.A.}    & \cthead{Dec.}   & \cthead{LSR radial} &
\cthead{$3$-$\sigma$} & \multicolumn{2}{c}{Distance}\\
                 & \cthead{1950}    & \cthead{1950}   &\cthead{velocity}    &
\cthead{(Jy)}       & \cthead{Near}&\cthead{Far}\\
                 & \cthead{(h m s)} & \cthead{(${}^\circ \quad {}\arcmin \quad {
} \arcsec$)} & \cthead{(km s$^{-1}$)} & & \cthead{(kpc)} & \cthead{(kpc)}\\
\hline

175.84$-$21.48   &  04:16:08.98  &      19:17:54.9  &     0.00 &  7.4 & 0.1 &
 17.1 \\
176.23$-$20.87   &  04:19:06.98  &      19:25:44.9  &     0.00 &  8.0 & 0.1 &
 17.1 \\
N105             &  05:10:12.64  &   $-$68:57:07.8  &   250.70 &  1.6 &&\\
210.42$-$19.76   &  05:33:53.17  &   $-$06:46:49.9  &    10.00 &  9.0 & 1.2 &
 15.8 \\
205.45$-$14.55   &  05:43:35.87  &   $-$00:09:25.9  &    10.00 &  7.1 & 1.3 &
 16.7 \\
213.83+0.62      &  06:52:43.57  &   $-$00:27:11.0  &    10.00 &  8.3 & 1.1 &
 15.2 \\
232.62+1.00      &  07:29:54.97  &   $-$16:51:47.0  &    23.00 & 10.6 & 1.1 &
 15.2 \\
233.76$-$0.19    &  07:27:52.67  &   $-$18:26:08.0  &    43.00 &  5.9 & 4.2 &
 14.3 \\
243.16+0.37      &  07:50:18.37  &   $-$26:17:53.0  &    52.00 &  6.5 & 5.3 &
 13.0 \\
254.66+0.21      &  08:18:54.06  &   $-$36:03:00.0  &    64.00 &  5.8 & 7.2 &
 11.7 \\
IRAS08337$-$4028 &  08:33:42.40  &   $-$40:28:01.8  &  $-$2.00 &  5.2 & 0.3 &
  3.2 \\
263.25+0.52      &  08:47:00.46  &   $-$42:43:15.0  &    13.00 &  6.5 & 2.7 &
  4.7 \\
IRAS08546$-$4254 &  08:54:36.20  &   $-$42:54:06.0  &     9.00 &  2.8 & 2.3 &
  4.0 \\
IRAS09015$-$4843 &  09:01:33.20  &   $-$48:43:25.4  &    56.00 &  4.4 & 8.0 &
      \\
IRAS09018$-$4816 &  09:01:50.30  &   $-$48:15:57.9  &    16.00 &  2.7 & 3.7 &
  4.0 \\
IRAS09149$-$4743 &  09:14:54.10  &   $-$47:43:13.0  &     4.00 &  3.6 & 2.2 &
      \\
284.35$-$0.42    &  10:22:20.00  &   $-$57:37:25.0  &     7.00 &  3.0 & 1.4 &
  5.6 \\
285.32$-$0.03    &  10:30:04.23  &   $-$57:47:50.1  &     0.60 &  2.9 & 0.5 &
  5.0 \\
IRAS10303$-$5746 &  10:30:21.36  &   $-$57:46:44.0  &     1.00 &  3.7 & 0.6 &
  5.1 \\
IRAS10460$-$5811 &  10:46:06.30  &   $-$58:11:12.4  &  $-$2.00 &  3.0 & 0.1 &
  5.2 \\
IRAS10555$-$6242 &  10:55:34.80  &   $-$62:42:49.8  & $-$16.00 &  2.6 & 2.4 &
  3.5 \\
IRAS11097$-$6102 &  11:09:46.70  &   $-$61:02:06.0  & $-$30.00 &  3.1 & 2.9 &
      \\
291.58$-$0.43    &  11:12:58.10  &   $-$60:53:40.0  &    11.00 &  3.8 & 1.5 &
  7.8 \\
293.84$-$0.78    &  11:29:49.60  &   $-$61:58:12.5  &    36.90 &  3.5 & 3.9 &
 10.7 \\
293.95$-$0.91    &  11:30:22.70  &   $-$62:07:22.2  &    41.40 &  3.9 & 4.3 &
 11.2 \\
IRAS11332$-$6258 &  11:33:15.00  &   $-$62:58:13.0  & $-$10.00 &  3.4 & 0.8 &
  6.3 \\
298.22$-$0.33    &  12:07:16.90  &   $-$62:33:01.0  &    31.80 &  3.8 & 3.2 &
 11.2 \\
299.01+0.13      &  12:14:42.90  &   $-$62:12:21.0  &    19.00 &  3.8 & 2.0 &
 10.2 \\
IRAS12272$-$6240 &  12:27:15.90  &   $-$62:40:25.0  &     8.00 &  5.6 & 1.0 &
  9.6 \\
IRAS13080$-$6229 &  13:08:05.54  &   $-$62:29:58.1  & $-$33.00 &  3.7 & 2.9 &
  6.9 \\
IRAS13111$-$6228 &  13:11:07.49  &   $-$62:28:31.7  & $-$37.00 &  4.8 & 3.3 &
  6.5 \\
308.92+0.12      &  13:39:35.40  &   $-$61:53:47.0  & $-$30.00 &  2.7 & 2.3 &
 8.4 \\
IRAS13504$-$6151 &  13:50:27.70  &   $-$61:51:37.3  &     3.00 &  2.9 & 0.4 &
 11.4 \\
311.96+0.14      &  14:04:19.70  &   $-$61:08:34.0  & $-$30.00 &  3.8 & 2.1 &
  9.6 \\
IRAS14050$-$6056 &  14:05:05.40  &   $-$60:56:29.0  & $-$50.00 &  2.8 & 4.0 &
  7.4 \\
IRAS14159$-$6038 &  14:16:00.50  &   $-$60:38:00.0  & $-$10.00 &  2.5 & 0.6 &
 11.1 \\
318.05+0.09      &  14:49:52.91  &   $-$58:56:47.1  & $-$49.51 &  3.2 & 3.4 &
  9.2 \\
318.94$-$0.20    &  14:57:01.50  &   $-$58:47:14.2  & $-$36.10 &  4.1 & 2.5 &
 10.4 \\
322.17+0.62      &  15:14:50.20  &   $-$56:27:51.1  & $-$53.86 &  3.4 & 3.6 &
  9.8 \\
326.66+0.57      &  15:40:59.00  &   $-$53:57:24.1  & $-$41.23 &  3.4 & 2.8 &
 11.4 \\
326.662+0.521    &  15:41:12.87  &   $-$53:59:39.4  & $-$41.00 &  4.8 & 2.8 &
 11.4 \\
327.590$-$0.094  &  15:48:45.38  &   $-$53:54:22.8  & $-$86.30 &  2.4 & 5.4 &
  8.9 \\
327.945$-$0.115  &  15:50:42.65  &   $-$53:41:55.7  & $-$51.70 &  2.8 & 3.4 &
 11.0 \\
328.20$-$0.58    &  15:54:02.31  &   $-$53:53:37.1  & $-$40.79 &  3.5 & 2.8 &
 11.7 \\
329.339+0.148    &  15:56:43.74  &   $-$52:36:13.6  &$-$106.50 &  2.5 & 7.3 &
 \\
329.622+0.138    &  15:58:11.08  &   $-$52:25:38.2  & $-$60.10 &  3.2 & 3.9 &
 10.8 \\
329.610+0.114    &  15:58:13.81  &   $-$52:27:12.8  & $-$60.10 &  3.2 & 3.9 &
 10.8 \\
331.425+0.264    &  16:06:22.14  &   $-$51:08:14.9  & $-$88.60 &  3.2 & 5.4 &
 9.6 \\
331.120$-$0.118  &  16:06:34.50  &   $-$51:37:31.3  & $-$65.00 &  3.0 & 4.2 &
 10.7 \\
332.094$-$0.421  &  16:12:28.14  &   $-$51:10:59.6  & $-$61.40 &  3.1 & 4.0 &
 11.0 \\
332.351$-$0.436  &  16:13:43.46  &   $-$51:01:00.6  & $-$53.10 &  2.6 & 3.6 &
 11.5 \\
332.560$-$0.148  &  16:13:25.00  &   $-$50:39:50.1  & $-$51.00 &  2.7 & 3.5 &
 11.6 \\
335.78+0.17      &  16:26:03.00  &   $-$48:09:44.1  & $-$48.53 &  3.5 & 3.5 &
 12.0 \\
339.88$-$1.26    &  16:48:24.76  &   $-$46:03:34.0  & $-$31.87 &  3.5 & 2.8 &
 13.2 \\
345.41$-$0.94    &  17:06:02.01  &   $-$41:31:44.0  & $-$22.86 &  2.6 & 2.6 &
 13.9 \\
348.18$-$0.49    &  17:08:39.01  &   $-$38:27:06.0  &  $-$7.18 &  3.1 & 1.1 &
  15.9 \\
IRAS17424$-$2859 &  17:42:29.00  &   $-$28:59:20.0  &    28.80 &  3.3 & &\\
IRAS17432$-$2855 &  17:43:16.00  &   $-$28:55:05.0  &    51.90 &  2.2 & &\\
IRAS17433$-$2841 &  17:43:21.00  &   $-$28:41:15.0  &    34.80 &  2.9 & &\\
\hline
\end{tabular}
\end{minipage}
\end{table*}

\begin{table*}
  \begin{minipage}{140mm}
    \contcaption{}
    \begin{tabular}{lrrrrrr}
      \hline
\cthead{Source}  & \cthead{R.A.}    & \cthead{Dec.}   & \cthead{LSR radial} &
\cthead{$3$-$\sigma$} & \multicolumn{2}{c}{Distance}\\
                 & \cthead{1950}    & \cthead{1950}   &\cthead{velocity}    &
\cthead{(Jy)}       & \cthead{Near}&\cthead{Far}\\
                 & \cthead{(h m s)} & \cthead{(${}^\circ \quad {}\arcmin \quad {
} \arcsec$)} & \cthead{(km s$^{-1}$)} & & \cthead{(kpc)} & \cthead{(kpc)}\\
\hline

IRAS17470$-$2853 &  17:47:04.00  &   $-$28:53:13.0  &    17.30 &  2.6 & 7.4 &
  9.6 \\
25.53+0.38       &  18:33:51.38  &   $-$06:26:59.4  &    95.60 &  2.7 & 5.7 &
  9.7 \\
25.41+0.09       &  18:34:37.30  &   $-$06:41:42.7  &    97.40 &  2.5 & 5.7 &
  9.6 \\
25.48+0.06       &  18:34:53.26  &   $-$06:38:34.8  &    95.50 &  2.8 & 5.7 &
  9.7 \\
25.82$-$0.18     &  18:36:21.78  &   $-$06:27:24.2  &    91.70 &  3.1 & 5.5 &
  9.8 \\
26.57$-$0.25     &  18:38:00.13  &   $-$05:49:14.4  &   103.80 &  3.3 & 6.1 &
  9.1 \\
29.98$-$0.04     &  18:43:35.00  &   $-$02:42:19.0  &   102.60 &  3.0 & 6.4 &
  8.3 \\
IRAS18449$-$0115 &  18:44:59.00  &   $-$01:15:59.0  &    97.80 &  6.4 & 6.3 &
  8.3 \\
IRAS18469$-$0132 &  18:46:59.00  &   $-$01:32:38.0  &    84.60 &  2.6 & 5.3 &
  9.2 \\ \hline
\end{tabular}
\end{minipage}
\end{table*}

\begin{table*}
\begin{minipage}{115mm}
\caption{Comparison between Mopra and Onsala observations of
95 GHz sources.}
\label{Onsala_Mopra95}
\begin{tabular}{lrrrrrr}
\hline
\cthead{Source}&\multicolumn{3}{c}{Mopra}&\multicolumn{3}{c}{Onsala}\\
\cthead{name}&\cthead{LSR radial}&\cthead{Line}&\cthead{Flux}&\cthead{LSR radial}&\cthead{Line}&\cthead{Flux}\\
             &\cthead{velocity}&\cthead{width}&\cthead{density}&\cthead{velocity}&\cthead{width}&\cthead{density}\\
&\cthead{(km~s$^{-1}$)}&\cthead{(km~s$^{-1}$)}&\cthead{(Jy)}&\cthead{(km~s$^{-1}$)}&\cthead{(km~s$^{-1}$)}&
\cthead{(Jy)}\\
\hline
OMC-2    & 11.4 & 0.3 & 138(7.0)  & 11.4 & 1.1 & 69.9(1.1) \\
NGC2264  &  7.3 & 0.2 & 99.2(5.7) &  7.7 & 1.0 & 46.2(1.7) \\
W33MetC  & 32.7 & 0.5 & 21.2(3.1) & 32.3 & 1.6 & 12.6(1.8) \\
L379IRS3 & 18.0 & 2.7 & 10.1(1.2)  & 18.2 & 3.9 & 21.3(1.3) \\
W51e1/e2 & 55.9 & 5.3 & 17.7(0.5) & 55.9 & 9.5 & 21.8(2.0) \\
W51Met2  & 56.7 & 0.6 & 19.7(2.6) & 56.5 & 1.2 &  9.4(0.9) \\
\hline
\end{tabular}
\end{minipage}
\end{table*}

\vskip 1cm

\subsection{Comments on individual sources}

\begin{table*}
\begin{minipage}{115mm}
\caption{Comparison between Mopra and Parkes observations at 44 GHz sources.}
\label{MopraP}
\begin{tabular}{lrrrrrr}
\hline
\cthead{Source}&\multicolumn{3}{c}{Mopra}&\multicolumn{3}{c}{Parkes}\\
\cthead{name}&\cthead{LSR radial}&\cthead{Line}&\cthead{Flux}&\cthead{LSR radial
}&\cthead{Line}&\cthead{Flux}\\
             &\cthead{velocity}&\cthead{width}&\cthead{density}&\cthead{velocity
}&\cthead{width}&\cthead{density}\\
&\cthead{(km~s$^{-1}$)}&\cthead{(km~s$^{-1}$)}&\cthead{(Jy)}&\cthead{(km~s$^{-1}
$)}&\cthead{(km~s$^{-1}$)}&
\cthead{(Jy)}\\
\hline
269.20$-$1.13 &     9.4 & 0.7 & 15.1(2.3) &     9.4 & 0.3 & 23.2(1.6) \\
270.26+0.84   &     9.3 & 0.6 &  7.0(3.1) &     9.7 & 0.3 & 37.0(6.3) \\
294.97$-$1.73 &  $-$8.2 & 1.3 &  5.9(0.8) & $-$8.25 & 0.9 & 12.2(2.0) \\
305.21+0.21   & $-$42.3 & 0.6 & 26.1(2.0) & $-$42.3 & 0.5 & 75.7(3.7) \\
305.36+0.20   & $-$33.4 & 0.9 & 17.4(1.4) & $-$33.1 & 0.6 &  106(8.9) \\
316.76$-$0.02 & $-$39.4 & 0.7 & 12.9(1.6) & $-$39.3 & 0.4 & 26.9(2.2) \\
320.28$-$0.31 & $-$66.3 & 0.7 & 14.5(1.9) & $-$66.3 & 0.9 & 45.4(3.2) \\
323.74$-$0.27 & $-$49.8 & 2.9 &  4.9(0.4) & $-$49.9 & 0.4 & 23.1(4.2) \\
324.72+0.34   & $-$50.6 & 2.5 &  3.1(0.5) & $-$51.8 & 0.4 & 43.8(3.7) \\
327.29$-$0.58 & $-$44.6 & 2.5 & 14.3(1.1) & $-$44.3 & 0.9 & 26.1(3.1) \\
328.81+0.64   & $-$40.9 & 3.1 &  7.2(0.9) & $-$40.6 & 1.2 & 23.5(1.6) \\
328.24$-$0.55 & $-$41.2 & 1.8 &  7.3(0.4) & $-$41.1 & 0.3 & 11.8(2.2) \\
329.03$-$0.21 & $-$43.7 & 1.3 & 19.3(1.3) & $-$43.7 & 0.7 & 24.9(3.5) \\
331.13$-$0.25 & $-$91.0 & 1.0 & 20.5(0.7) & $-$91.0 & 0.7 & 95.0(4.4) \\
331.34$-$0.35 & $-$65.7 & 0.7 & 15.8(1.3) & $-$65.7 & 0.4 & 24.0(5.5) \\
333.23$-$0.05 & $-$87.2 & 1.3 & 99.3(2.3) & $-$87.2 & 1.3 &  205(6.6) \\
333.13$-$0.43 & $-$48.6 & 1.0 & 17.0(2.1) & $-$48.2 & 0.8 & 98.5(3.8) \\
333.61$-$0.22 & $-$49.7 & 2.9 &  3.0(0.4) & $-$49.3 & 0.6 & 21.2(2.5) \\
335.59$-$0.29 & $-$45.5 & 0.6 & 22.3(2.2) & $-$45.4 & 0.6 &  158(4.5) \\
336.41$-$0.26 & $-$87.6 & 0.7 & 24.8(0.9) & $-$87.6 & 0.7 & 51.4(3.7) \\
337.40$-$0.41 & $-$42.3 & 0.2 & 16.5(2.4) & $-$42.3 & 0.2 & 26.8(2.3) \\
338.92+0.56   & $-$62.7 & 1.8 & 43.3(1.9) & $-$62.9 & 0.5 &  166(5.4) \\
337.91$-$0.47 & $-$43.6 & 0.4 & 58.2(2.4) & $-$43.4 & 0.4 &  273(4.0) \\
341.19$-$0.22 & $-$42.0 & 0.7 &  6.1(0.9) & $-$41.9 & 0.3 & 92.9(2.9) \\
341.22$-$0.21 & $-$43.8 & 1.8 & 15.6(1.0) & $-$43.8 & 0.4 & 25.5(6.8) \\
345.01+1.79   & $-$13.1 & 0.5 & 49.6(1.9) & $-$13.2 & 0.4 & 33.4(2.7) \\
343.12$-$0.06 & $-$32.8 & 0.8 & 46.6(3.0) & $-$32.8 & 1.0 & 71.9(2.2) \\
344.23$-$0.57 & $-$20.3 & 0.6 & 33.2(1.9) & $-$20.3 & 0.6 & 48.4(9.6) \\
345.51+0.35   & $-$16.5 & 0.9 &  3.2(0.5) & $-$16.5 & 0.6 & 16.2(1.6) \\
345.00$-$0.22 & $-$28.0 & 0.7 & 23.5(1.4) & $-$27.9 & 0.6 & 69.3(9.4) \\
349.10+0.11   & $-$78.0 & 0.5 &  6.4(1.6) & $-$77.9 & 0.3 & 23.8(2.1) \\
351.16+0.70   &  $-$4.9 & 0.6 & 26.3(3.3) &  $-$4.5 & 0.4 & 36.9(9.9) \\
351.24+0.67   &  $-$4.5 & 3.4 &  4.3(0.4) &  $-$4.5 & 0.3 & 21.5(0.8) \\
351.41+0.64   &  $-$8.2 & 4.2 & 20.2(1.0) &  $-$8.4 & 1.4 & 39.9(1.9) \\
351.78$-$0.54 &  $-$6.9 & 0.4 & 73.0(2.3) &  $-$6.9 & 0.4 &  150(4.6) \\
351.64$-$1.26 & $-$12.7 & 0.7 & 15.2(2.5) & $-$13.1 & 1.2 & 12.7(1.1) \\
354.61+0.47   & $-$17.9 & 0.5 & 24.0(1.7) & $-$17.8 & 0.5 & 55.5(9.8) \\
353.41$-$0.36 & $-$16.6 & 0.6 &  8.1(0.8) & $-$16.5 & 0.6 & 31.4(3.5) \\
359.62$-$0.25 &    19.3 & 0.6 & 15.9(1.1) &    19.4 & 0.4 &  105(5.2) \\
0.54$-$0.85   &    16.9 & 0.6 &  4.0(0.9) &    14.8 & 0.7 & 33.8(1.7) \\
M8E           &    11.0 & 0.5 &129.7(2.0) &    10.9 & 0.4 &  510(34.2) \\
12.89+0.49    &    31.5 & 0.4 &  5.7(1.1) &    31.4 & 0.3 & 24.4(2.2) \\
14.33$-$0.64  &    23.4 & 0.5 & 52.8(2.9) &    23.5 & 0.5 &  120(9.8) \\
16.59$-$0.06  &    61.3 & 0.8 &  9.8(1.6) &    61.2 & 0.2 & 12.5(2.5) \\
23.43$-$0.19  &    99.6 & 0.5 & 33.6(2.1) &    99.7 & 0.4 & 78.9(6.1) \\
23.01$-$0.41  &    77.4 & 1.3 &  3.4(0.6) &    77.1 & 0.3 & 32.2(2.8) \\
30.69$-$0.06  &    89.3 & 0.6 &  3.8(0.6) &    89.5 & 0.5 & 29.8(7.7) \\
\hline
\end{tabular}
\end{minipage}
\end{table*}

{\bf OMC-2}.  The 95-GHz methanol maser in this source consists of a
very intense narrow line similar to that observed at 44~GHz (Haschick
et al. 1990).  The peak flux densities at both frequencies are almost
equal, as are the line widths.

{\bf NGC2264}. This source shows an intense narrow line at 95~GHz with
a shoulder on the positive velocity side.  The spectrum of the 95-GHz
methanol maser is very similar to that at 44-GHz, but the peak flux
density is a factor of two lower.

{\bf 305.21+0.21}. The spectra of the 95- and 44-GHz masers are very
similar, showing a single intense narrow line with the 95-GHz flux density
only 25\% lower than the 44-GHz flux density.

\begin{figure}
\vspace{5.3cm}\includegraphics{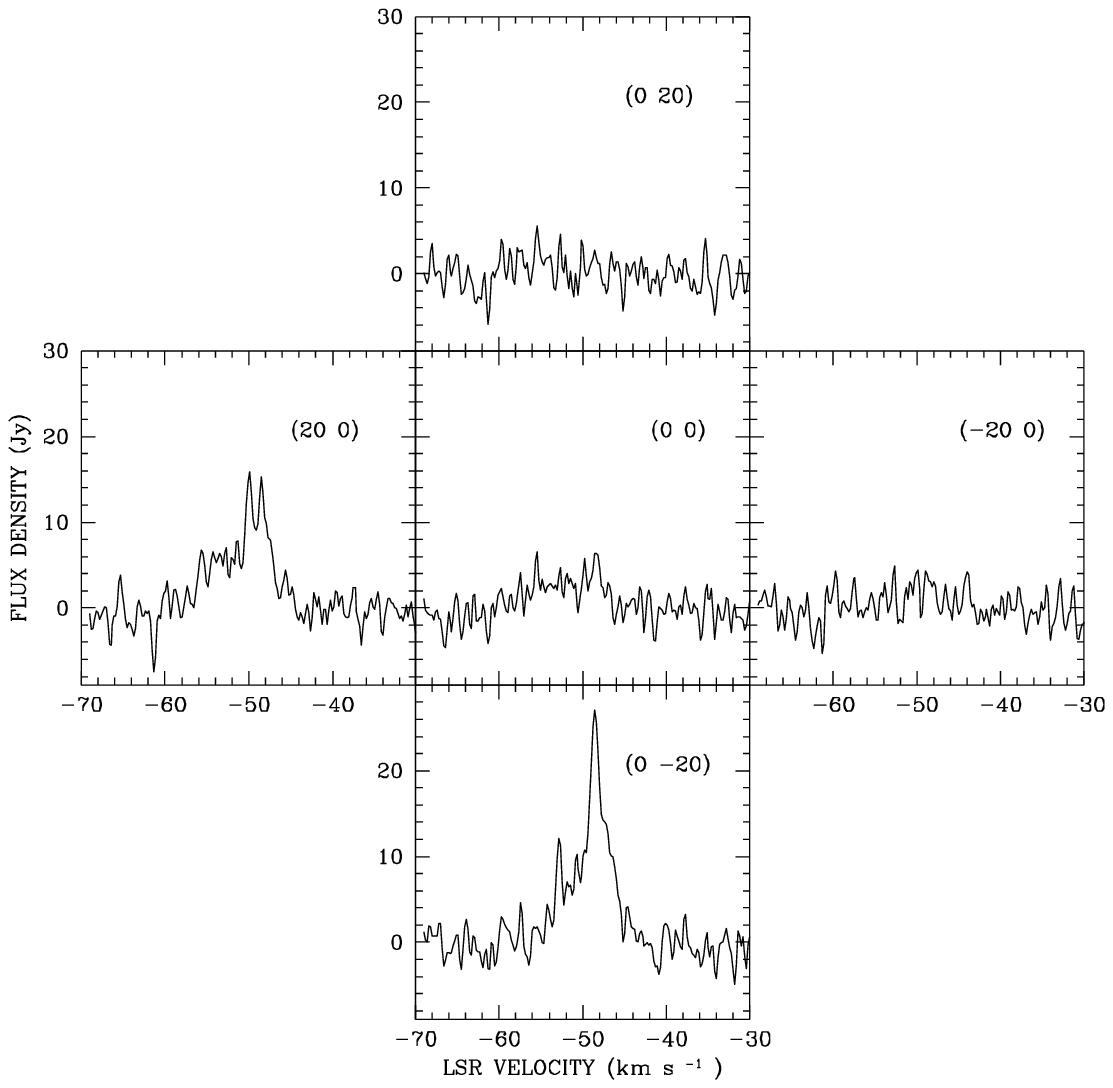}
\caption{Map of 333.13-0.43.}
\label{333_13}
\end{figure}

{\bf 333.13-0.43}.  The spectrum of this source in Fig.1 shows a weak
broad line.  Comparing this spectrum with the 44-GHz spectrum we
suspected that 95-GHz flux density was too low, possibly because of a
pointing error. To check this we made a five-point map consisting of
an observations at the central position and offset by $\pm 20''$ in
both right ascension and declination.  Fig.~\ref{333_13} contains the
spectra from this map, which shows that the strongest emission is
offset to the south-west of the nominal position and that there are at
least two rather strong narrow lines confirming that this is a maser.


\begin{figure}
\vspace{5.3cm}\includegraphics{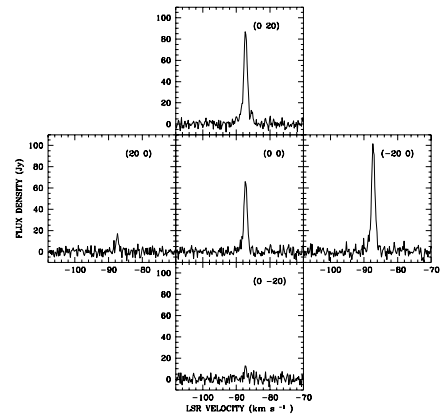}
\caption{Map of 333.23-0.05.}
\label{333_23}
\end{figure}

{\bf 333.23-0.05}. This source (Fig.1) has an intense isolated
spectral feature, which is best fitted by two Gaussians.  A
five-point map (Fig.~\ref{333_23}) shows that the source is in fact
offset from the nominal position by approximately $-$20$''$ in right
ascension.  The 95-GHz maser spectrum is similar to that at 44~GHz,
where it is a factor of 3 stronger.

{\bf 335.59-0.29}. The 95-GHz spectrum of this source is best fitted with
4 Gaussian components, while at 44~GHz only one intense component is
present at the same radial velocity as the strongest 95-GHz component.

{\bf 337.91-0.47}. The 95-GHz methanol maser spectrum is essentially
identical to the 44-GHz spectrum, but the peak flux density at 44~GHz
is a factor of four higher.

\begin{figure}
\vspace{5.3cm}\includegraphics{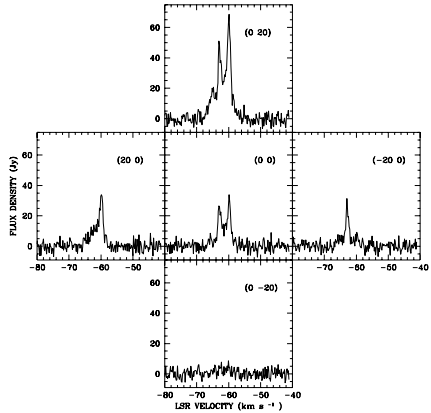}
\caption{Map of 338.92$+$0.56.}
\label{338_92}
\end{figure}

{\bf 338.92+0.56}.  The 95-GHz spectrum of this source has two narrow
features, at radial velocities $-$62.9~\ks and $-$60.1~\ks, the same
as the 44-GHz spectrum (Slysh et al. 1994).  However, a five-point map
centred on the nominal position of the source (Fig.~\ref{338_92})
shows that the strongest emission is to the north, where the flux
density is at least a factor of two higher. From the spectrum of the
offset (0, 20$''$) one can see also a third detail, at the radial
velocity $-$65.2~\ks, which is not present in the 44-GHz spectrum.
The five-point map also shows that the position of the spectral
feature at $-$60.1~\ks is shifted in right ascension relative to the
spectral feature at $-$62.9~\ks by about 20$''$.

\begin{figure}
\vspace{5.3cm}\includegraphics{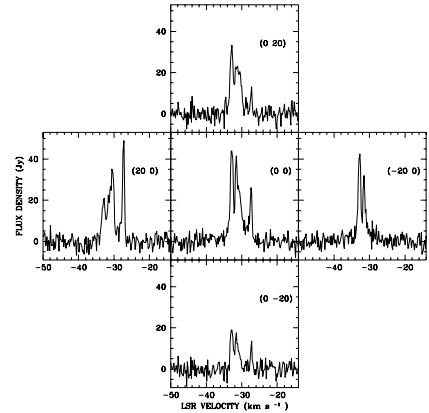}
\caption{Map of 343.12-0.06.}
\label{343_12}
\end{figure}

{\bf 343.12-0.06}. The 95-GHz spectrum of this source has been fitted
with four Gaussian components, all of which have counterparts in the
44-GHz spectrum (Slysh et al. 1994), although the relative intensities
are different at the two frequencies.  A five-point map of the source
(Fig.~\ref{343_12}) shows that the emission in the source is spread
over an area of at least 20$''$, so the difference in relative
intensities is most likely due to the different telescope beamwidths
and pointing errors between the 44- and 95-GHz observations.

{\bf 345.01+1.79} The 95-GHz spectrum of this source has been fitted
by two Gaussians separated by 1.2~\ks.  A 9-point map of this source
(Val'tts 1998) was used to determined position of both components. The
stronger component at -13.1~\ks has a position which coincides within
the errors with the position of the southern 6.7-GHz methanol maser
(Norris et al. 1993). The weaker component at $-$14.3~\ks is displaced
from the stronger one by $-$6.6~$\pm$7.9$''$ in right ascension and
10.2$\pm$5.3$''$ in declination, and coincides within the errors with
the position of the northern 6.7-GHz methanol maser. Thus 95-GHz
methanol masers are present in both regions of methanol maser emission
in this source.

{\bf 345.00-0.22}.  The 95-GHz methanol maser spectrum is similar to
that at 44~GHz, except for the presence of an additional feature at
$-$24~\ks, which is absent from the 44-GHz spectrum.

{\bf 351.16+0.70 (NGC6334B)}. The 95-GHz spectrum is similar to 44-GHz
spectrum, although due to blending of spectral features the number of
Gaussian components we are able to fit is different, 4 at 95~GHz
compared to 7 at 44~GHz.

{\bf 351.24+0.67 (NGC6334C)}. The 95-GHz emission from this source
is anomalously low compared to 44-GHz emission (Slysh et al. 1994),
possibly due to poor pointing, or variable weather conditions.

{\bf 351.41+0.64 (NGC6334F)} A broad line with some weaker narrow
features is present in the 95-GHz spectrum; we have fitted the profile
with two broad components and one narrow at the radial velocity
$-$6.4~\ks. It is probably the counterpart of the -6.32~\ks component
in the 44-GHz spectrum (Slysh et al. 1994), but other narrow 44-GHz
features are not present in the 95-GHz spectrum.

\begin{figure}
  \vspace{9cm}\includegraphics{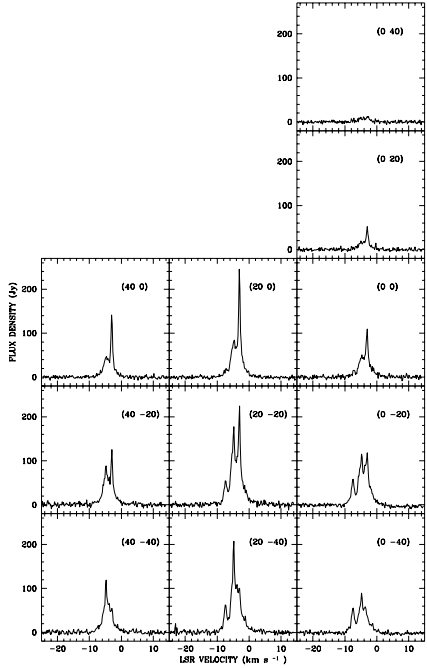}
\caption{Map of NGC6334I(N).}
\label{ngc6334i}
\end{figure}

{\bf NGC6334I(N)}.  This is one of the strongest methanol masers at
both 95- and 44-GHz (Haschick et al. 1990). The spectra are similar at
the two frequencies, considering likely differences in
pointing, since the maser is known to be spread over an area more than
30$''$ in extent as can be seen from our map (Fig.~\ref{ngc6334i}).
The relative positions of the four strongest spectral features were
determined from this map and are given in Table~\ref{t4}.  A
comparison with the map of the source at 44~GHz (Kogan \& Slysh 1998)
shows that there is a general agreement between the two maps if larger
position errors of 95-GHz map are taken into account. All four
spectral features are point-like within the Mopra beamwidth of 52$''$.

\begin{figure}
  \vspace{5.3cm}\includegraphics{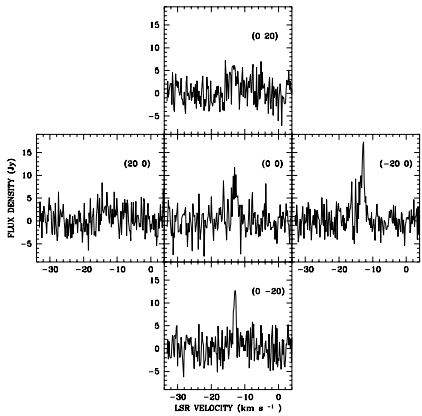}
\caption{Map of 351.64-1.26.}
\label{351_64}
\end{figure}

{\bf 351.64$-$1.26}.  A five-point map of this source shows that the
single weak feature becomes stronger at offsets of ($-$20$''$,0) and
(0,$-$20$''$) (Fig.~\ref{351_64}), which implies that the true source
position is to the south-east of the the nominal position.

{\bf 351.78-0.54}.  The 95-GHz spectrum consists of four narrow
components, all of which are present in the 44-GHz spectrum, and in
addition there is a broad component about 8~\ks wide, which is absent
at 44~GHz.

{\bf 354.61+0.47}.  At least three Gaussian components can be fitted
to the 95-GHz spectrum of this source, and two of them are present in
the 44-GHz spectrum.  The component at $-$21.4~\ks is relatively
stronger at 95~GHz and the component at $-$20.5~\ks is present only at
95~GHz.

\begin{table}
\caption{Position of components in NGC6334I(N)}
\label{t4}
\begin{tabular}{cccrcr}
\hline
\cthead{N}&\cthead{LSR radial}&\cthead{Line}&\cthead{$\Delta
\alpha$}&\cthead{$\Delta \delta$}&\cthead{Flux}\\
          &\cthead{velocity}&\cthead{width}&&&\cthead{density}\\

&\cthead{(km~s$^{-1}$)}&\cthead{(km~s$^{-1}$)}&($\prime\prime$)&($\prime\prime$)
&\cthead{(Jy)}\\
\hline

1 & $-$7.5 & 0.8 &  9(5) & $-$30(5) &  44 \\
2 & $-$4.8 & 0.8 & 18(1) & $-$35(5) & 110 \\
3 & $-$4.0 & 4.0 & 18(5) & $-$28(5) & 100 \\
4 & $-$3.0 & 0.5 & 24(6) & $-$12(7) & 175 \\
\hline
\end{tabular}
\end{table}

\begin{figure}
  \vspace{9cm}\includegraphics{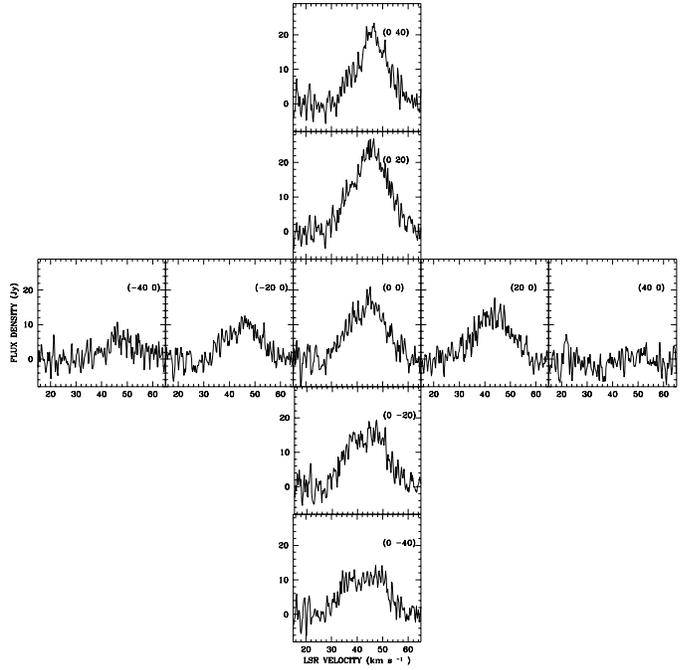}
\caption{Map of Sgr A-A.}
\label{sgraa}
\end{figure}

{\bf Sgr A-A}.  The 95-GHz spectrum of this source consists of a very
broad component and possibly a weak narrow component at a velocity of
47~\ks.  It is almost identical to the 44-GHz spectrum (Haschick et
al. 1990), with the narrow component relatively stronger at 44~GHz.  A
9-point map (Fig.~\ref{sgraa}) of the source does not show the narrow
component, probably because of the shorter integration time of map
spectra. The broad component is extended in declination with an
angular size of 150$''$, and is unresolved (less than 30$''$) in right
ascension. This is one of the rare strong thermal sources in our
sample.

{\bf M8E}.  A single very strong narrow line is present in the 95-GHz
spectrum, which is similar to that at 44~GHz. This is one of the
strongest masers in both the 44- and 95-GHz transitions, and its
angular extent is known to be less than 0.2$''$ (Kogan \& Slysh 1998,
Slysh et al. 1999). A thorough discussion of the methanol emission
from this source is given in Val'tts (1999a).

{\bf W33MetC}.  A single feature near 32.5~\ks is present in 95-GHz
spectrum.  At 44~GHz there is also a weaker component at 36.28~\ks
which is not visible in 95-GHz spectrum. The 95-GHz emission in this
source has been mapped by Pratap \& Menten (1992), and the
corresponding 44-GHz maser has been shown to coincide with its
position (Slysh et al.  1999).  A full discussion of the methanol
emission from this source is given by Val'tts (1999b).

{\bf 14.33-0.64}.  This intense class~I methanol maser was discovered
at 44~GHz at Parkes (Slysh et al. 1994).  At 95-GHz the maser is also
strong and its spectrum consists of four peaks, similar to the 44-GHz
spectrum.  A 44-GHz map of this maser is discussed in Slysh et al.
(1999).

\begin{figure}
  \vspace{9cm}\includegraphics{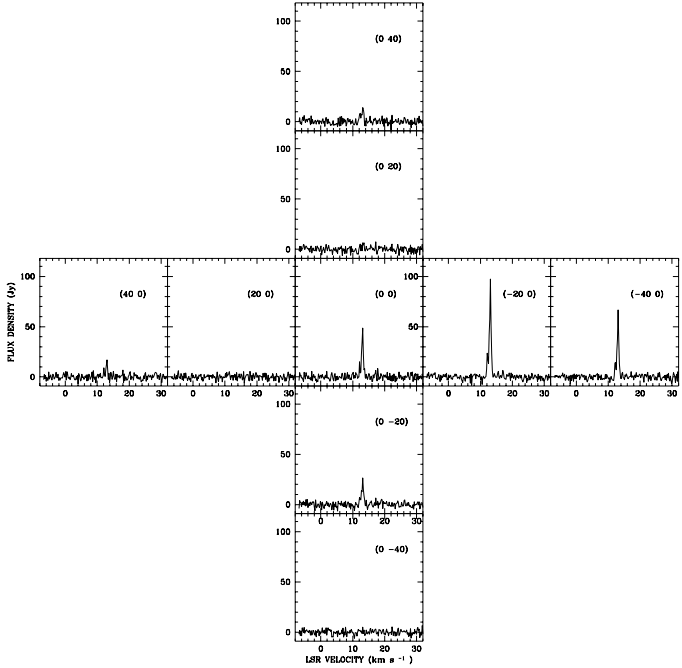}
\caption{Map of GGD27.}
\label{ggd27}
\end{figure}

{\bf GGD27}. 44-GHz class~I methanol maser emission in this source was
discovered by Kalenskii et al. (1992), and a VLA map of it is
presented in Slysh et al. (1999). At both 95 and 44~GHz the spectrum
is dominated by a strong narrow feature.  A 9-point map of the 95-GHz
emission (Fig.~\ref{ggd27}) shows that it is strongest at a position
offset in right ascension by $-$23$''$$\pm$8$''$ and in declination by
$-$8$''$$\pm$5$''$ from the nominal position.

\begin{figure}
  \vspace{5.3cm}\includegraphics{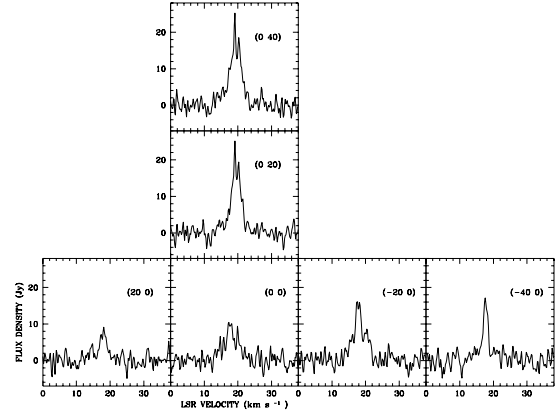}
\caption{Map of L379IRS3.}
\label{l379}
\end{figure}

{\bf L379IRS3}. The 44-GHz methanol maser associated with this source
was also discovered by Kalenskii et al. (1992) and mapped with the VLA
by Kogan \& Slysh (1998) and Slysh et al.  (1999).  The 95-GHz
spectrum is very similar to that at 44~GHz and a map (Fig.~\ref{l379})
of the 95-GHz emission shows that the four spectral features are
spread over an area of approximately 30$''$, similar to the 44-GHz
emission.

{\bf 23.43-0.19}. Two Gaussian components have been fitted to the
detection at 95~GHz of this source, both are present in the
44-GHz spectrum (Slysh et al. 1994). In addition there are several
components near 103~\ks in the 44-GHz spectrum that are not present at
95~GHz.

{\bf IRAS18537+0749 (S76E)}.  This rather strong source was discovered
to be a class~I methanol maser during observations of the
$4_{-1}-3_0E$ transition at 36~GHz at Puschino (Val'tts, private
communication).  In both transitions it appears that a blend of
several narrow lines is producing a spectrum resembling a wide band of
emission.

{\bf W51e1/e2}. Only a broad component is present in our 95-GHz
spectrum.  The narrow spectral feature at 48.88~\ks detected by
Haschick et al. (1990) at 44~GHz is not visible in the 95-GHz
spectrum, because it is shifted from W51e1/e2 by about 70$''$ and was
outside the main beam of Mopra telescope (Pratap \& Menten 1992).

\section{Discussion}

The spectra of the 95-GHz $8_0-7_1A^+$ methanol emission sources found
in this survey are in general similar to the spectra of the
corresponding 44-GHz $7_0-6_1A^+$ sources.  The emission in the two
transitions typically covers the same velocity range, has
approximately the same number of spectral features with very similar
radial velocities, and in some cases even the same relative
intensities of the components. In Table~\ref{MopraP} we list the single
strongest feature in each spectrum at 95~GHz and the corresponding
spectral features at 44~GHz of the sources detected by Slysh et al.
(1994).  One can see that there is always a corresponding spectral
feature at 44~GHz to every 95~GHz spectral feature from
Table~\ref{t4}, and their radial velocities agree in general to within
0.1~\ks. The line width of the 95-GHz components is in general
somewhat larger than the line width of the corresponding 44-GHz
features, partly due to a lower spectral resolution in the 95~GHz
observations, but nevertheless there are many very narrow 95-GHz
features with a line width less than 1~\ks.  The peak flux density of
the 95-GHz components is generally lower than the flux density of the
44-GHz features.  Fig.~\ref{MopraOnsala} shows a comparison between
the flux densities of spectral features with the same radial
velocities from the two transitions. In constructing this plot data on
all available sources were used, including the results of this work
and of the observations at Onsala (Val'tts et al. 1995).  The straight
line (with a correlation coefficient $r$=0.73) shows the best fit
linear dependence which was found to be :

\begin{figure*}
\resizebox{\hsize}{!}{\includegraphics{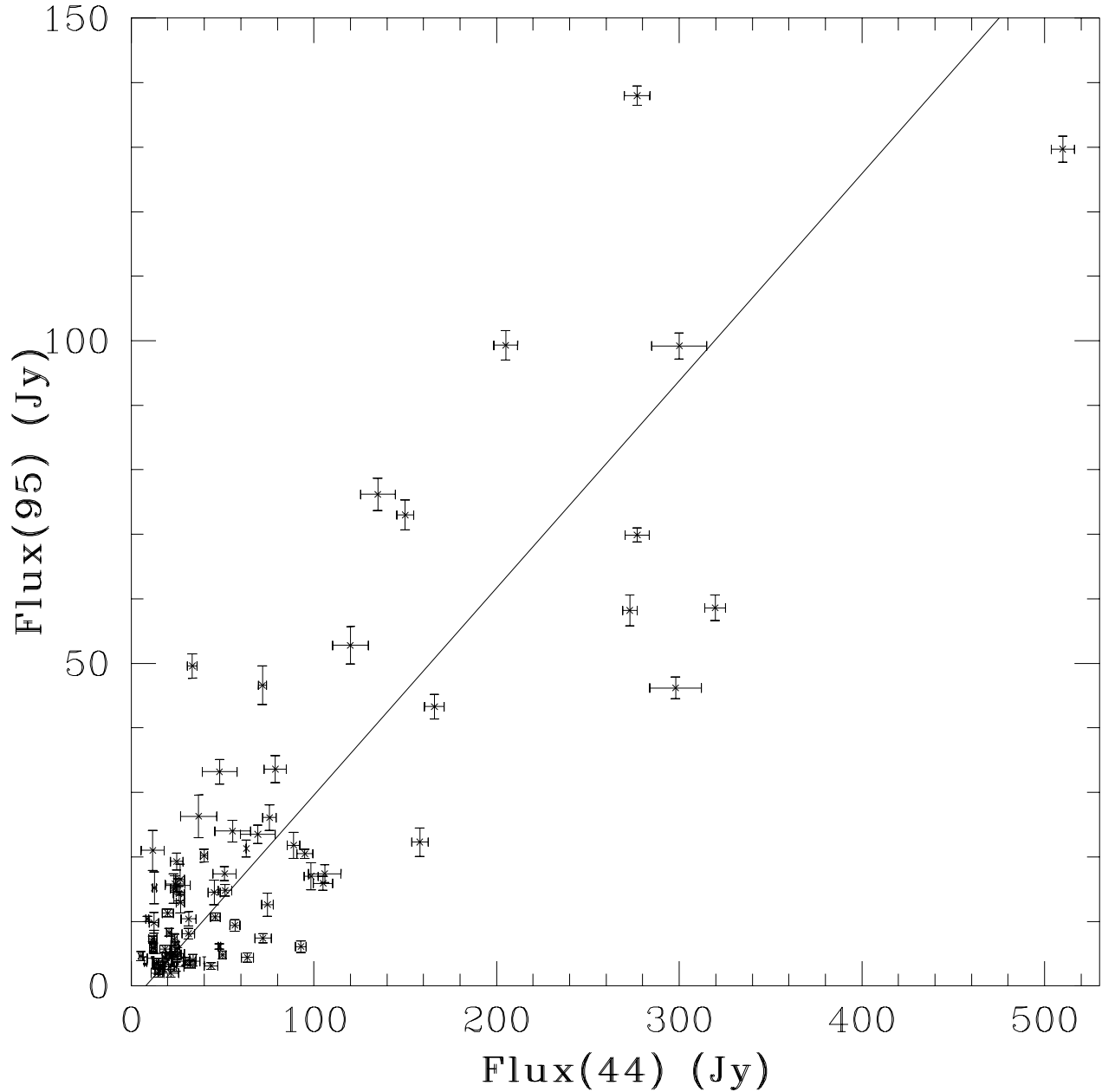}}
\caption{A correlation between 95 GHz and 44 GHz peak flux density. Straight
line is a best fit: $y$=(0.32$\pm$0.08)$x$-(8.1 $\pm$ 2.7), correlation coefficient - $r$=0.73. }
\label{MopraOnsala}
\end{figure*}

\begin{equation}
S(95)=(0.32 \pm 0.08) \times S(44) - (8.1 \pm 2.7)
\end{equation}

Although the scatter is quite large, on average the 95-GHz methanol
masers are a factor of 3 weaker than the 44-GHz masers.  This result
is consistent with the findings of Val'tts et al. (1995) who found a
linear dependence between integrated fluxes at two frequencies.  We
have used peak flux densities rather than luminosities since our
survey is flux density limited by the sensitivity of the instrument,
and luminosities would give a spurious correlation due to the
multiplication of the flux densities at the two frequencies by the
same distance squared.  The slope of the dependence between integrated
flux densities found by Val'tts et al. (1995) is $0.52\pm0.05$, which
is larger than the slope $0.32\pm0.08$ found in this paper for the
peak flux densities. This difference may be due to the larger average
line width of 95-GHz masers mentioned above.  The correlation between
the peak flux density and the observed similarity in the spectra of
the two transitions is strong evidence in favour of the suggestion
that the emission from both transitions arises from the same spatial
location.  A comparison of published high resolution maps of the 44-
and 95-GHz class I methanol masers in DR21(OH) and W33MetC shows that
their images are very alike and consist of the same number of isolated
maser spots (Plambeck \& Menten 1990, Pratap \& Menten 1992, Kogan \&
Slysh 1998, Slysh et al. 1999), consistent with this hypothesis.

The two transitions belong to the class~I methanol masers (Menten
1991), which are thought to be pumped through collisional excitation.
The difference between the two transitions is that the upper level of
the 95-GHz $8_0-7_1A^+$ transition is 18.5~K above the upper level of
the 44-GHz $7_0-6_1A^+$ transition.  Therefore the population of the
former is expected to be lower than the population of the latter,
resulting in the lower intensity of 95-GHz emission compared to the
intensity of the 44-GHz transition, although it is difficult to
estimate the difference without any knowledge of the kinetic
temperature and particle density in the source.

\begin{table}
\caption{LVG calculation results: ratios of 44/95 intensities for four models.
Methanol density, divided by velocity
gradient, for all models is $0.67\times 10^{-2}$ cm$^{-3}$/(km s$^{-1}$/pc).}
\label{LVGtable}
\begin{tabular}{lrrr}
\hline
\cthead{Model}&\cthead{$T_{kin}$}&\cthead{n$_{H_2}$}&\cthead{44/95 GHz}\\
              &   (K)            &\cthead{}         &\cthead{intensity ratio}\\
\hline
1             & 20               & 0.56E+5          & 3.3 \\
2             & 50               & 0.56E+5          & 1.7 \\
3             & 100              & 0.56E+5          & 0.4 \\
4             & 20               & 0.56E+6          & 1.2 \\
\hline
\end{tabular}
\end{table}

We used LVG code to calculated the intensity ratios of the
$7_0-6_1A^+$ and $8_0-7_1A^+$ transitions in a collisional excitation
model for four different parameter sets. The model parameters and the
intensity ratios are presented in Table~\ref{LVGtable}.  The
collisional selection rules are based on the paper by Lees \& Haque
(1974) and imply that $\Delta K=0$ collisions are preferred by a
factor of four.  For model 1 with a gas temperature 20~K and density
$0.56\times 10^5$ cm$^{-3}$, the ratio of the 44- and 95-GHz
intensities is 3.3, i.e., close to the mean observed ratio. The 95~GHz
intensity is lower due to the lower population of the $8_0A^+$ level
relative to that of the $7_0A^+$ level and due to a weaker inversion
at 95~GHz.  Increasing either the gas temperature or the density
decreased the model ratio below the observed value.  Thus, our results
favour class I maser model with gas temperature about 20 K and density
less than 10$^6$ cm$^{-3}$.

\section{Summary}
\begin{enumerate}
\renewcommand{\theenumi}{\arabic{enumi}.}
\item As a result of a survey in the southern hemisphere 85 methanol
  emission sources were detected in the $8_0-7_1A^+$ transition at
  95~GHz. This survey together with a similar Onsala survey (Val'tts
  et al. 1995) completes a whole sky survey of methanol emission at
  95~GHz.

\item Most of the detected sources are class~I methanol masers, and
  the majority of them have counterparts in other class~I methanol
  transitions, such as the $7_0-6_1A^+$ at 44~GHz.

\item The previously found correlation between the methanol maser emission
  intensity at 44 and 95~GHz is confirmed here, using a larger sample
  of sources.

\item A maser model with collisional excitation based on LVG
  calculations can explain the observed intensity ratio at 44 and
  95~GHz and gives constraints on the temperature and particle
  density.

\end{enumerate}

\section{Acknowledgements}

I.E.V. is grateful to the ATNF for the hospitality, and to the staff
of Mopra observatory for the help with the observations.  The
Australia Telescope is funded by the Commonwealth of Australia for
operation as a National Facility managed by CSIRO.  Travel to
Australia for I.E.V. was aided by grant 96/1990 from the Australian
Department of Industry, Science and Tourism.  The work of I.E.V.,
V.I.S., S.V.K. and G.M.L. was partly supported by the grants
95-02-05826 and 98-02-16916 from the Russian Foundation for Basic
Research and by the Federal Program "Astronomiya" (Project N 1.3.4.2).
S.P.E thanks the Queen's trust for the computing system used to
process the data from these observations.  The authors would like to
thank Ms V. Oakley and Mr J. Saab for their assistance during the
observations and initial data processing.

\label{lastpage}

\end{document}